\documentclass[amsmath,amssymb,onecolumn,superscriptaddress]{revtex4}

\usepackage{graphicx}
\usepackage{dcolumn}

\def\be{\begin{equation}}
\def\ee{\end{equation}}
\def\beq{\begin{eqnarray}}
\def\eeq{\end{eqnarray}}

\usepackage{bm}
\usepackage{graphicx}
\usepackage{dcolumn}
\usepackage{amsmath}
\usepackage[latin1]{inputenc}
\usepackage{graphicx, psfrag}
\usepackage{amssymb}
\usepackage[colorlinks=true, citecolor=blue, urlcolor = blue, linkcolor= red, bookmarks=true]{hyperref}
\usepackage{float}
\usepackage{amsmath}
\usepackage{tensor}
\usepackage{amsfonts}
\usepackage{dcolumn}
\usepackage{hyperref}
\usepackage{subfigure}
\usepackage{pgfplots}
\usepackage{epstopdf}
\usepackage{booktabs}

\begin{document}


\title{Probing black hole in Starobinsky-Bel-Robinson gravity 
 with thermodynamical analysis, effective force and gravitational weak lensing} 

\author{G. Mustafa}
\email{gmustafa3828@gmail.com}
\affiliation{Department of Physics, Zhejiang Normal University, Jinhua 321004, People's Republic of China}
\affiliation{Institute of Fundamental and Applied Research, National Research University TIIAME, Kori Niyoziy 39, Tashkent 100000, Uzbekistan}

\author{Allah Ditta}%
\email{mradshahid01@gmail.com}
\affiliation{Department of Mathematics, Shanghai University and Newtouch Center for Mathematics of Shanghai University,  Shanghai,200444, P.R.China.}

\author{Faisal Javed}
\email{faisaljaved.math@gmail.com} 
\affiliation{Department of Physics, Zhejiang Normal University, Jinhua 321004, People's Republic of China}
\affiliation{Institute of Fundamental and Applied Research, National Research University TIIAME, Kori Niyoziy 39, Tashkent 100000, Uzbekistan}

\author{Farruh Atamurotov}
\email{atamurotov@yahoo.com}
\affiliation{New Uzbekistan University, Movarounnahr street 1, Tashkent 100000, Uzbekistan}
\affiliation{Central Asian University, Milliy Bog' Street 264, Tashkent 111221, Uzbekistan}
\affiliation{University of Tashkent for Applied Sciences, Str. Gavhar 1, Tashkent 100149, Uzbekistan}

\author{Ibrar Hussain}
\email{ibrar.hussain@seecs.nust.edu.pk}	
\affiliation{School of Electrical Engineering and Computer Science, National University of Sciences and Technology, H-12, Islamabad, Pakistan}

\author{Bobomurat Ahmedov}
\email{ahmedov@astrin.uz}

\affiliation{Institute of Fundamental and Applied Research, National Research University TIIAME, Kori Niyoziy 39, Tashkent 100000, Uzbekistan} \affiliation{Ulugh Beg Astronomical Institute, Astronomy St 33, Tashkent 100052, Uzbekistan} 
\affiliation{Institute of Theoretical Physics, National University of Uzbekistan, Tashkent 100174, Uzbekistan}

\begin{abstract}
In this work, we investigate the effects of plasma and the coupling parameter $\beta>0$, on the thermodynamic properties and weak gravitational lensing by the Schwarzschild-like black hole in the Starobinsky-Bel-Robinson gravity (SBRG). We observe that the horizon radius and the corrected entropy of the Schwarzschild-like black hole in the SBRG are not much sensitive to the parameter $\beta$. On contrary the energy emission rate of the Schwarzschild-like black hole in the SBRG is sensitive to the parameter $\beta$ and decreases with increase in the values of the parameter $\beta$. We see that the Schwarzschild-like black hoe in the SBRG is stable as the thermodynamicl temperature is positive for different values of the parameter $\beta$. Moreover we observe that the deflection angle of photon beam by the black hole in uniform plasma, nonuniform self-interacting scalar plasma and non-singular isothermal gas sphere reduces with the parameter $\beta$, against the impact parameter $b$. We see that the deflection angle enhances with increase in the concentration of the plasma fields for all the three types of plasma media. Further we find that the magnification of the image due to lensing increases in a higher concentration of plasma field. It is interesting to notice that the image magnification in uniform plasma is much higher as compared to the one in nonuniform plasma field. We compare our results with those for the Schwarzschild black hole of General Relativity. Further, effective force is also calculated for the current analysis. 
\end{abstract}

\date{\today}

\maketitle

\section{Introduction}

One of the most important topics in the present state of research, the study of black holes (BHs) is widely recognized. The most recognizable characteristics of powerful classical and quantum gravitational fields are displayed by such thermodynamical objects. On the basis of classical reasoning, BH contains extraordinary gravitational forces that prevent any kind of particle or radiation from crossing the event horizon and destroying everything in its surroundings. By supporting the quantum mechanical process, the BHs are able to generate and emit thermal radiation known as Hawking radiations \cite{1}. As a result of the BH mass being steadily reduced by these radiations, it eventually evaporates. BHs display Hawking temperatures and entropy that vary for various kinds of BHs since they are thermal objects. The relationship between classical thermodynamics and the BH rules may be seen in the temperature and entropy of BHs. More specifically, surface gravity is connected to temperature, whereas energy is related to BH mass. Entropy exhibits the relationship with the field of the event horizon of BHs  and has a significant role in assessing the thermal characteristics of a thermodynamic system \cite{2}. The entropy of BHs should be higher than that of any other object with a volume similar to BH in order to prevent breaking the second law of thermodynamics. To illustrate the requirement for logarithmic corrections based on thermal fluctuations in the entropy area relation shown by Bekenstein \cite{3}, it is difficult to reach thermal equilibrium between thermal radiations and BH.

It is an intriguing subject to investigate the influences of quantum oscillations on the geometrical properties of BHs. The fluctuations produced by statistical perturbations in compact celestial bodies like BHs, are recognized as thermal fluctuations, which may depict the marvelous features associated with the BH geometry. A majority of BHs are thought to shrink in size as a result of Hawking radiation, resulting in an increase in temperature of the BH. Pourhassan and Faizal \cite{3a} examined the influence of the thermal fluctuations on the thermodynamic potentials for a spinning Kerr-AdS BH and estimated the accuracy of the BH entropy using a logarithmic correction factor. The same results were obtained for higher-dimensional BHs corresponding to the higher-order correction terms \cite{3b,3c}. Pourhassan et al. \cite{3d} demonstrated the effects of logarithmic adjustments. By choosing the cosmological constant, Jawad and Shahzad examined the thermodynamic stability and thermal oscillations associated with non-minimal regular BHs. They discovered that the conventional BHs portray stability for a wider range of cosmological constants. Zhang \cite{3f} used Kerr-Newman-AdS and Reissner-Nordstrom BHs to explain the first-order correction terms to the entropy that affect the thermodynamic factors of smaller BHs. For three-dimensional Godel BH, the Hawking temperature and vector particle tunneling have also been explored \cite{3g}.  Pradhan \cite{3h} investigated the stability of thermal oscillations using charged BHs. Extensive studies have also been conducted in the literature to examine the phase transitions in addition to thermal fluctuations for different types of BHs \cite{3i,3j,3k,3l,3m,3n,3o,3p,3q}.

The deflection of light and production of a lens-like effect are the results of gravitational lensing, which is caused by the curvature of spacetime around massive compact objects. Gravitational lensing is used to test the validity of General Relativity theory and to investigate the characteristics of matter near BHs. According to theories of gravity, BHs contain powerful gravitational forces that prohibit anything from passing beyond the event horizon. It is critical that the gravitational lensing effect can distinguish between various BH lenses.  In the weak and strong field limits, gravitational lensing is still a very active research area \cite{Eiroa:2005ag,Wei:2011bm}. The strong gravitational lenzing of Schwarzschild BH is studied by Virbhadra et al. \cite{Virbhadra:1999nm}. A Schwarzschild lens simulation of the supermassive BH M87* is used in the investigation to examine how changes in the angular source position affect the images' tangential, radial, and overall magnification \cite{Virbhadra:2022iiy}. Under the weak field approximation, Sereno \cite{Sereno:2003nd} developed the formulas for the time delay and deflection angle of Reissner-Nordstrom BH. In many modified gravity theories, the weak deflection angle has been examined using a variety of techniques \cite{Jusufi:2017vta,Ovgun:2018oxk,Li:2020wvn,Fu:2021akc,Javed:2020pyz,Javed:2021arr,Li:2021xhy,Crisnejo:2019xtp,Crisnejo:2019ril,Jha:2021eww}. In general, derivatives of the elements of the BH metric can be used to define the deflection angle or corresponding optical scalar. Numerous publications have been written on the subject of gravitational lensing in strong gravity fields, which is a prominent research area \cite{Virbhadra:2002ju,Rahvar:2018nhx,Bozza:2010xqn,Virbhadra:2008ws,Chen:2013vja,Ji:2013xua,Chen:2015cpa,Chen:2016hil,Zhang:2017vap,Abbas:2019olp,Abbas:2021whh,Hensh:2021nsv}.

It is generally accepted that plasma makes up the challenging environment in which compact astrophysical objects exist. The effect of plasma on a spherically symmetric BH in the SBRG is especially investigated in this study. A dispersive medium like plasma has a refractive index that depends on photon frequency. Due to possible interactions with electromagnetic waves. The existence of plasma near compact astrophysical objects can change the trajectories of light rays. Synge \cite{Synge:1960ueh} was the first to suggest the self-consistent method for investigating the movement of light rays in a gravitational field within a plasma medium. Afterwards, Perlick \cite{Perlick2000} developed a new approach to determine the deflection angle created when plasma envelops Schwarzschild and Kerr BHs, yielding an integral formula. Further research by Bisnovatyi-Kogan and Tsupko \cite{Bisnovatyi-Kogan:2008qbk} revealed that the deflection angle in a uniformly dispersive medium, which is qualitatively distinct from the vacuum environment, depends on the photon frequency.

By developing formulas for several plasma models, the authors \cite{Bisnovatyi-Kogan:2010flt} studied the deflection angle in the presence of plasma inhomogeneities enveloping a gravitational object. Gravitational lensing around typical BHs submerged in plasma was studied by Schee et al. \cite{Schee:2017hof} . In a prior research \cite{Turimov:2022iff}, the weak deflection angle of a wormhole solution characterized by an exponential metric was established. In a non-minimally linked Einstein-Yang-Mills (EYM) theory, the effects of uniform plasma on the shadow and weak deflection angle were examined for spinning and regular BHs \cite{Kala:2022uog}. The shadow of the Kerr BH was studied by Zhang et al. in relation to plasma with a logarithmic normal distribution and a power-law distribution. Additionally, Atamurotov and his colleagues concentrated on studying the weak gravitational lensing effect in plasma for a number of different spacetimes, such as rotating Einstein-Born-Infeld BH \cite{Babar:2021nst}, the Lorentzian wormhole spacetime \cite{Atamurotov:2021byp}, 4D Einstein-Gauss-Bonnet gravity \cite{ad36}, Schwarzschild-MOG BH \cite{Atamurotov:2021qds}, and Schwarzschild BH with cloud of stings and other modified matter content \cite{rn1,rn2,rn3}.

BHs have recently been studied using modified gravity models, which are currently being employed to tackle some of the Universe's challenges. Numerous hypotheses have been put out utilizing various approaches and techniques. In particular, Einstein Gauss-Bonnet (EGB) gravity, which is supported by string theory and associated dual theories like M-theory, is the one that has received the most attention of the researchers \cite{r1,r2,r3}. By considering  a parameter referred as the Gauss-Bonnet gravity parameter, BHs in these gravity models have been constructed and studied \cite{r4,r5}. The development of the novel SBRG has been made possible by the addition of a new stringy parameter \cite{r6}. In one of recent works the impact of the parameter $\beta$ on the tidal force and shadow of a BH in the SBRG has been studied \cite{Himanshu2023}.In another recent work the rotating version of the SBRG BH has been studied in the context of of photon motion in the presence of plasma \cite{pdu23}. In the present work, we consider BH solution in SBRG to explore their more physical features. The paper is partitioned as follows:  Section II presents the BH solution in SBR gravity and investigate the thermodynamical configurations. Then, emission energy of the considered BH spacetime is studied in Section III. Section IV deals with the corrected entropy, and gravitational lensing with plasma is explored in Section V. Further, the magnification of gravitational lensed image is discussed in Section VI. In the last section, we present some concluding remarks.

\section{BLACK HOLES IN SBR GRAVITY}\label{II}

This section is devoted to explore the Schwarzschild-like BH solution in the SBRG which is theorized to be embedded in M-theory existing in eleven dimensional spacetime. This solution is discussed by the authors in \cite{ad46}. M-theory has a bosonic sector having a metric $g_{M N}$ and a tensor 3-form $C_{M N P}$ coupled to M2-branes which are dual to M5-branes \cite{ad51}. Using the compactification process coupled with the existence of stringy fluxes, the corresponding 4-D gravity models may be produced \cite{ad46,adn48,adn47}. The action considered is given by:
\begin{equation}
S_{S B R}=\frac{M_{p l}^{2}}{2} \int d^{4} x \sqrt{-g}\left(R+\frac{R^{2}}{6 m^{2}}-\frac{\beta}{32 m^{6}}\left(P_{4}^{2}-E_{4}^{2}\right)\right).  
\label{1}
\end{equation}
where $R$ is the Ricci curvature scalar and $g$ is the metric determinant. Here $m$ is a free mass parameter which could have various interpretations depending on the underlying theory. Also, $\beta$ is a positive coupling whose value is determined by the compactification of the M-theory, and could be fixed by studying the optical behaviors of the BH. Where $P_{4}^{2}$ and $E_{4}^{2}$ are the Pontryagin and the Euler topological densities which are related to the Bel-Robinson tensor $T_{\mu \nu \lambda \rho}$ in four dimensions by means of the relation \cite{ad46,adn48,adn47}

\begin{equation}
T^{\mu \nu \delta \rho} T_{\mu \nu \delta \rho}=\frac{1}{4}\left(P_{4}^{2}-E_{4}^{2}\right).  
\label{2}
\end{equation}

The Bel-Robinson tensor is defined as:

\begin{equation}
T^{\mu \nu \delta \sigma}=R^{\mu \rho \gamma \delta} R_{\rho \gamma}^{\nu \sigma}+R^{\mu \rho \gamma \sigma} R_{\rho \gamma}^{\nu \delta}-\frac{1}{2} g^{\mu \nu} R^{\rho \gamma \alpha \delta} R_{\rho \gamma \alpha}^{\sigma}.
\label{3}
\end{equation}
It can be seen from Eq.(\ref{1}) that the gravity action depends on two parameters ($m,\beta$). This gives rise to plethora of applications such as studying Hawking radiation, entropy  and inflation in the study of cosmology. The BH solution in SBRG depends on the value of $\beta$ corrected to first order perturbations. The line element of this non-rotating solution has been found to be \cite{adn47}

\begin{equation}
d s^{2}=-f(r) d t^{2}+\frac{1}{f(r)} d r^{2}+r^{2} d \Omega^{2}.
\label{4}
\end{equation}
where the metric component $f(r)$ is given by

\begin{equation}
f(r)=1-\frac{R_{s}}{r}+\beta\left(\frac{4 \sqrt{2} \pi  G R_{s} }{r^{3}}\right)^{3}\left(\frac{108 r- 97 R_{s}}{5 r}\right).
\label{5}
\end{equation}
\begin{figure}
\includegraphics[scale=0.30]{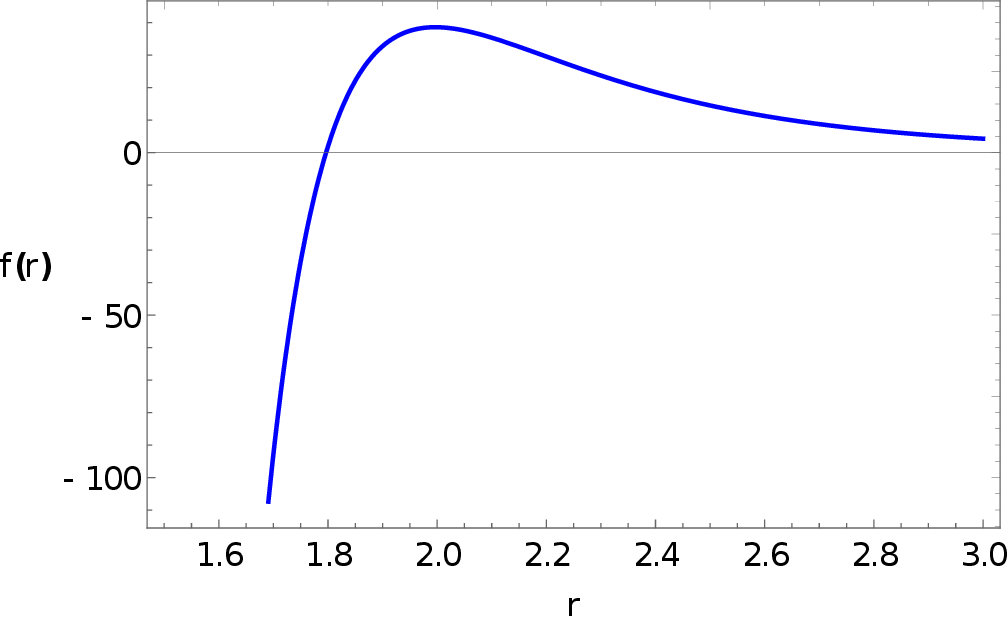}
\includegraphics[scale=0.30]{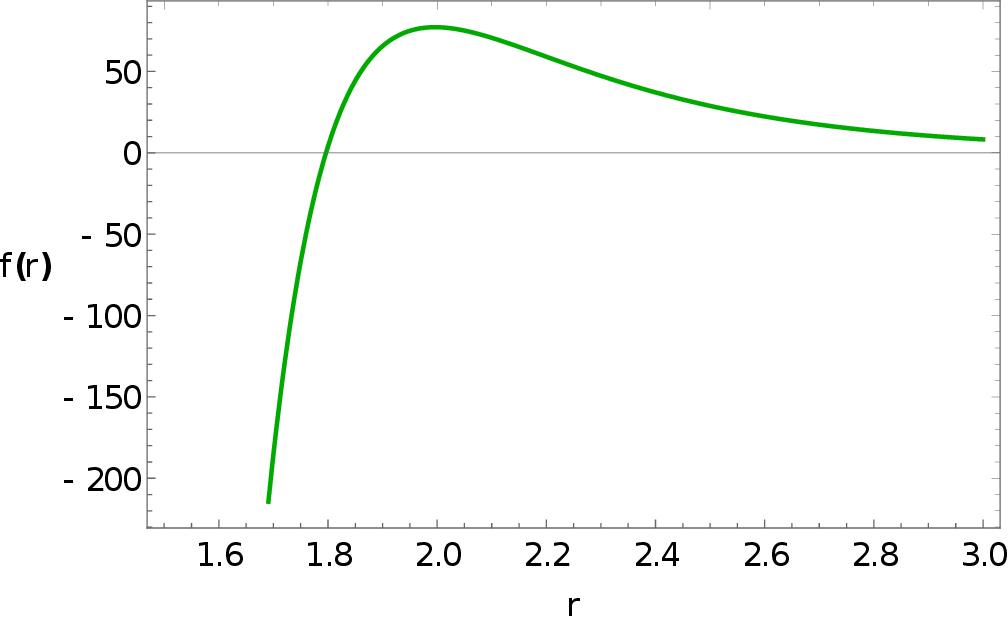}
\includegraphics[scale=0.30]{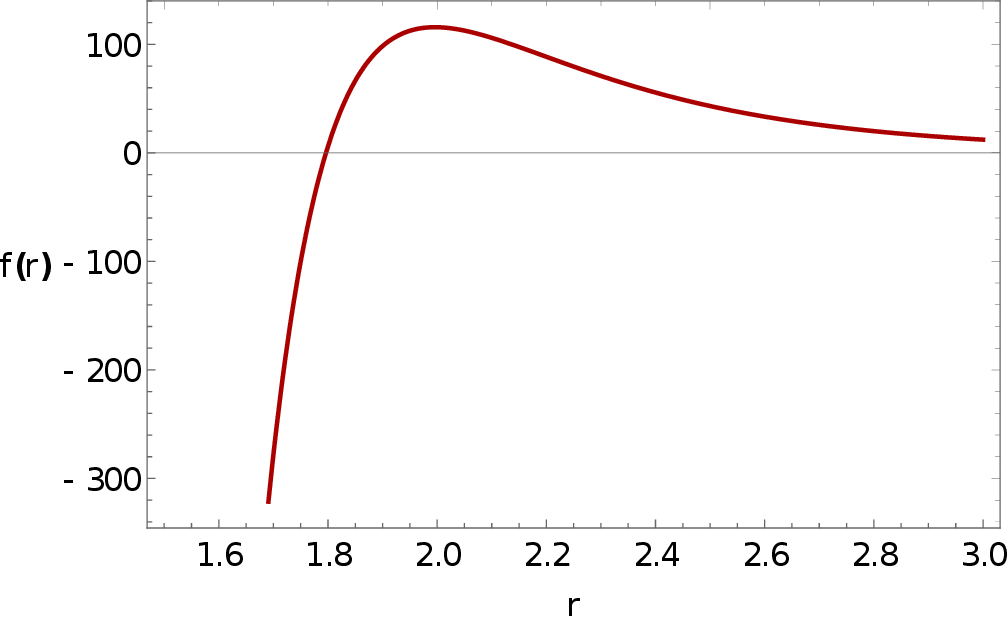}
    \caption{Lapse function $f(r)$ along the horizon radius $r_{0}$. Herein we have fixed values $M=1,\;G=1$ with $\beta=0.2$ (Left graph), $\beta=0.4$ (Middle graph), $\beta=0.6$ (Right graph).}\label{p1}
    \end{figure}
   
     \begin{figure}
   \includegraphics[scale=0.30]{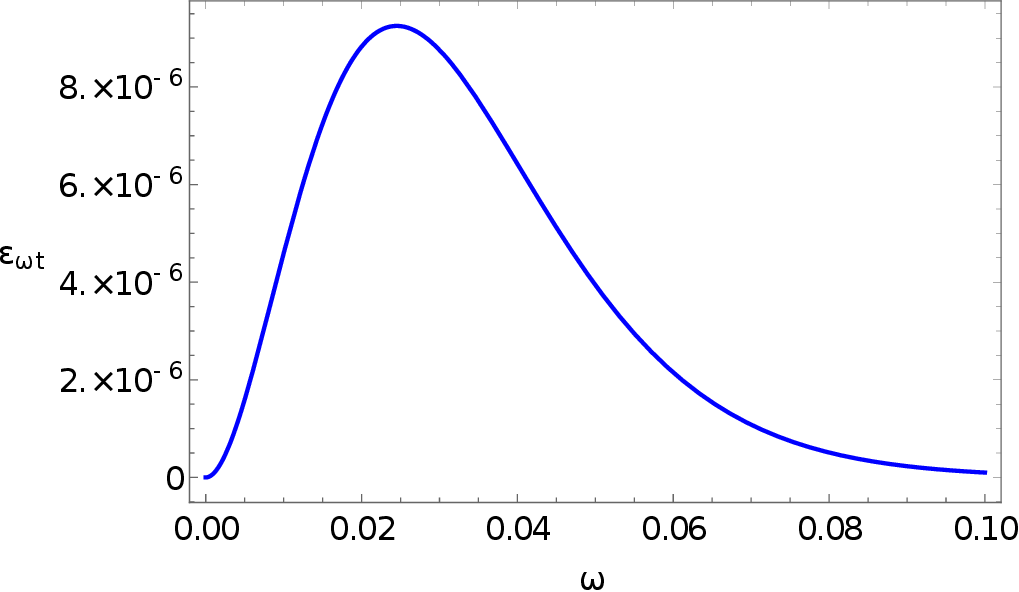}
   \includegraphics[scale=0.30]{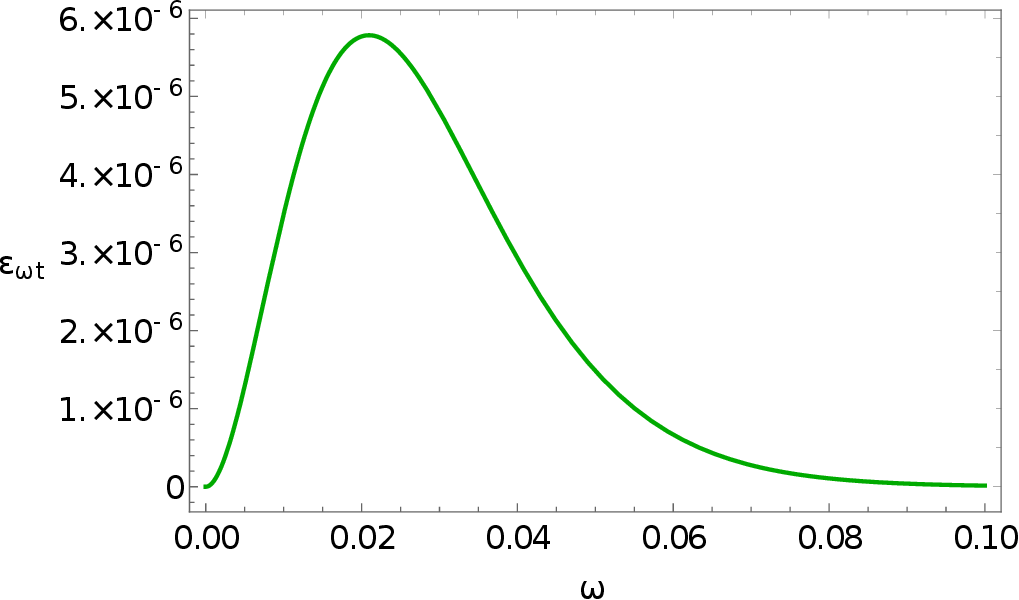}
    \includegraphics[scale=0.30]{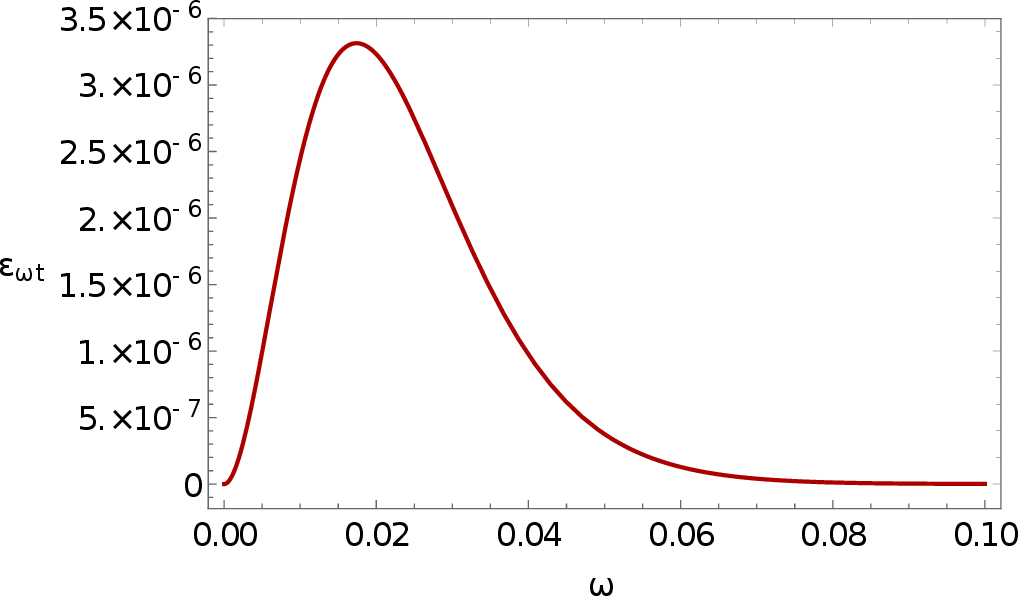}
    \caption{Energy emission along $\omega$ and along the horizon radius $r_0$. Herein we have fixed values $M=1,\;G=1$ with $\beta=0.2$ (Left graph), $\beta=0.4$ (Middle graph), $\beta=0.6$ (Right graph).}
    \label{plot:6}
    \end{figure}

Where $R_s=2GM$ is the Schwarzschild radius and $M$ is the mass of the BH. The possible existence of positive and negative event horizons can be predicted by the physical behavior of lapse function $f(r)$ as shown in Fig. (\ref{p1}). It is obvious that positive $\beta$ depict the positive event horizon and increasing values of $\beta$ depict the increasing possibility of positive event horizon.The Hawking temperature $T_{H}$ of BH can be determined by $\frac{f^{'}(r)}{4\pi}$ as:
\begin{equation}\label{9aa}
    T_H=\frac{993280 \pi ^3 \sqrt{2} \beta  G^7 M^4-497664 \sqrt{2} \pi ^3 \beta  G^6 M^3 r_{0}+5 G M r_{0}^{9}}{10 \pi  r_{0}^{11}}.
\end{equation}
The Entropy of the BH can be written as:
\begin{equation}
S=\pi r_0^2.
\end{equation}

\section{ Emission Energy}\label{E}
Quantum fluctuations occurring within the interior of BHs result in the continual creation and annihilation of particles beyond the event horizon in large quantities. This tunneling phenomenon \cite{ad15, ad16, ad17, ad18, ad19} causes positively charged particles to be attracted toward the innermost region of the BH, which leads to the emission of Hawking radiation and eventual evaporation of the BH over a specific time period. The rate of evaporation is directly proportional to the energy emission rate. From the perspective of a distant observer, the high-energy reception cross-section closely approximates the BH shadow. This energy reception cross-section exhibits oscillations around a fixed, constrained value denoted as $\sigma_{lim}$, which corresponds to the radius of the BH \cite{ad20,ad21,ad22,ad23,f1,f2}: 
\begin{equation}\label{9}
\sigma_{lim}\approx \pi r_{0}^{2},
\end{equation}
where, $r_{0}$ is the event horizon radius of the BH. Thus the expression for the energy emission rate of the BH is:
\begin{equation}\label{10}
\frac{d^{2}\varepsilon}{d\omega dt}=\frac{2\pi^{2}\sigma_{lim}}{e^{\frac{\omega}{T}}-1}\omega^{3}=\frac{\left(2 \pi ^3 r_{0}^{2}\right) \omega ^3}{\exp \left(\frac{10 \pi  r_0^{11}\omega }{993280 \pi ^3 \sqrt{2} \beta  G^7 M^4-497664 \sqrt{2} \pi ^3 \beta  G^6 M^3 r_0 + 5 G M r_{0}^{9}}\right)-1}
\end{equation}
One can obtain information about emission energy from Fig. \ref{plot:6}. One can see that $\varepsilon_{\omega t}$ increases by decreasing $\beta$.
\section{Corrected Entropy}
\begin{figure}
    \includegraphics[scale=0.12]{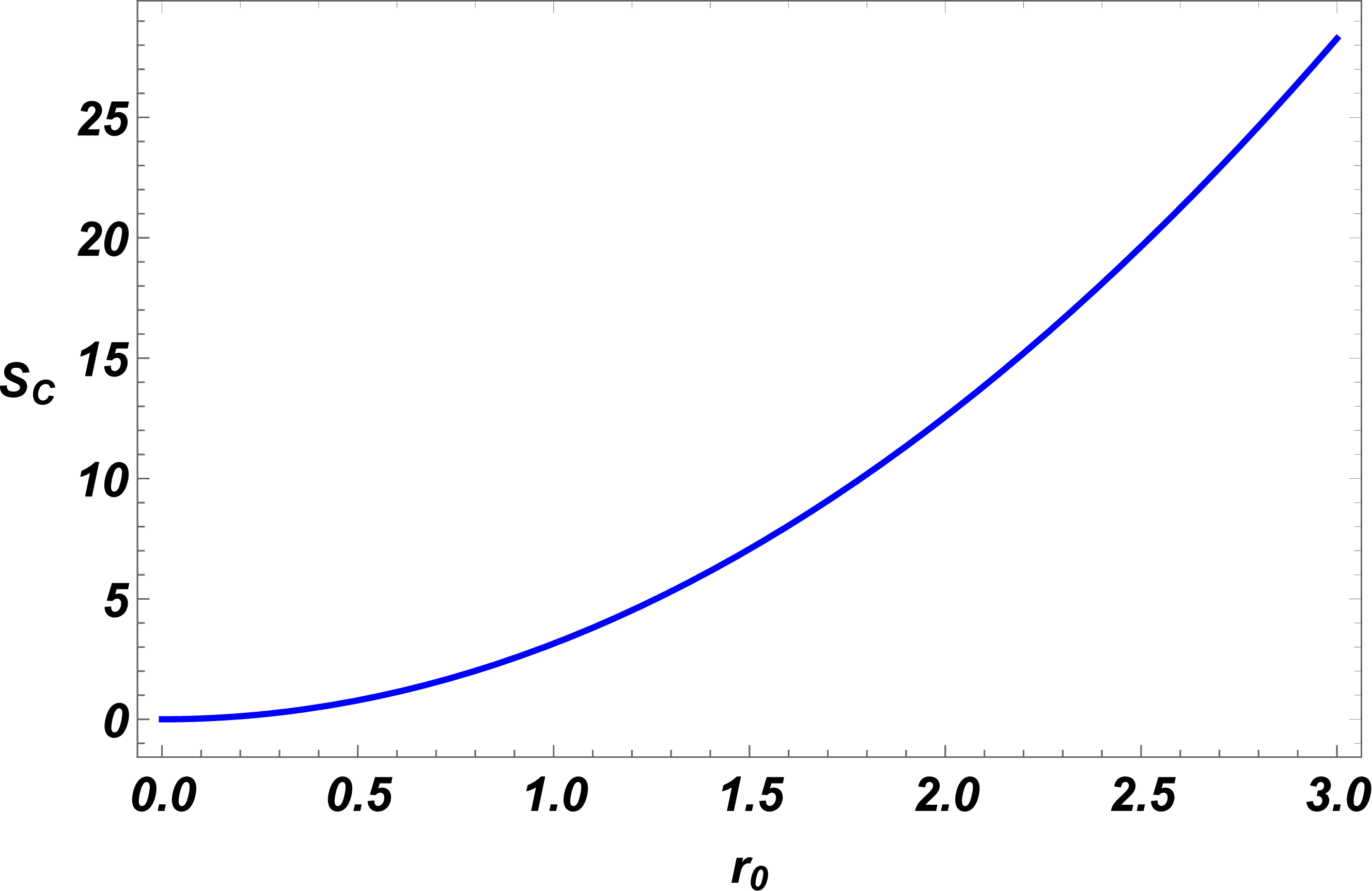}
     \includegraphics[scale=0.12]{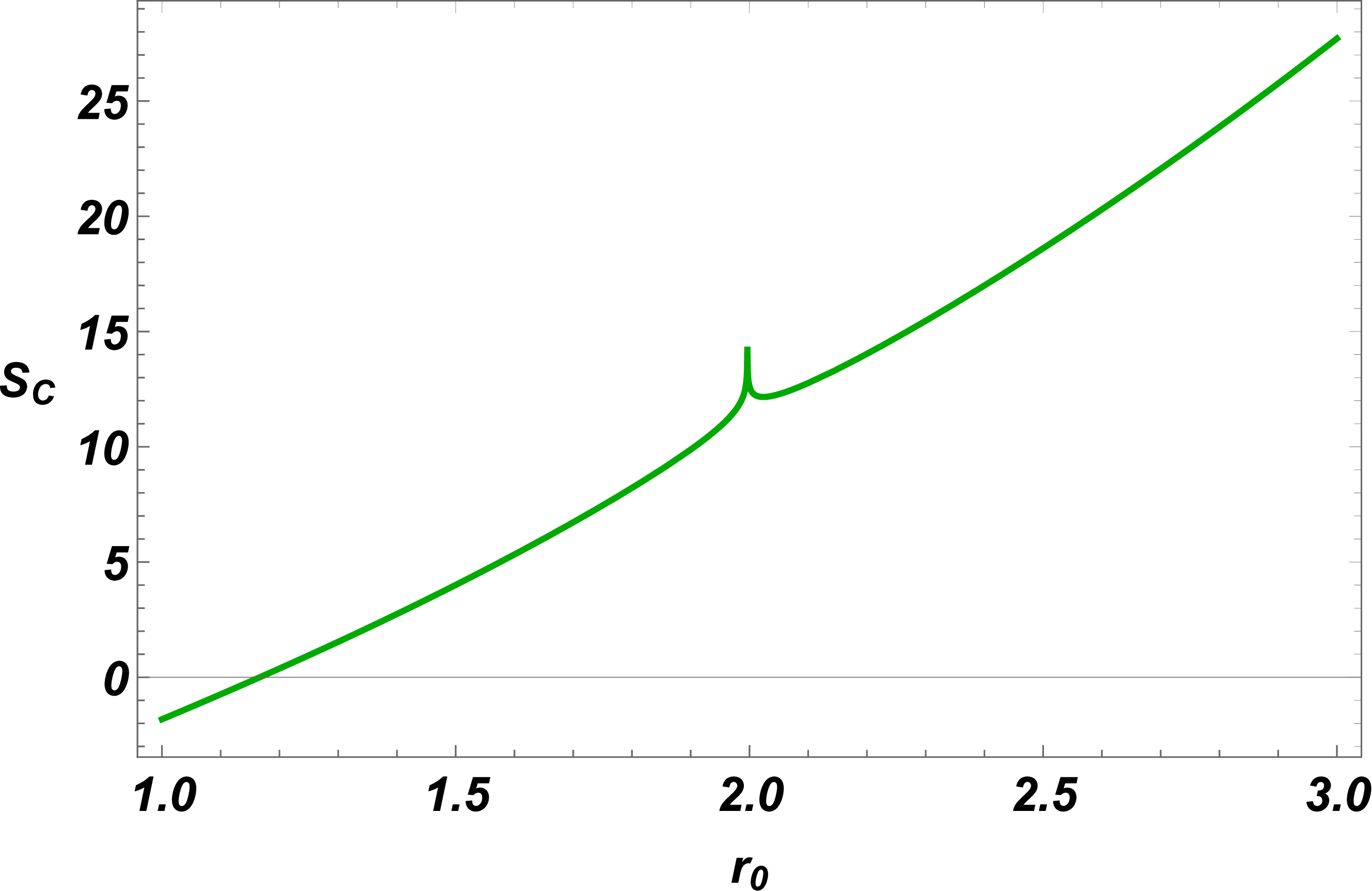}
    \includegraphics[scale=0.12]{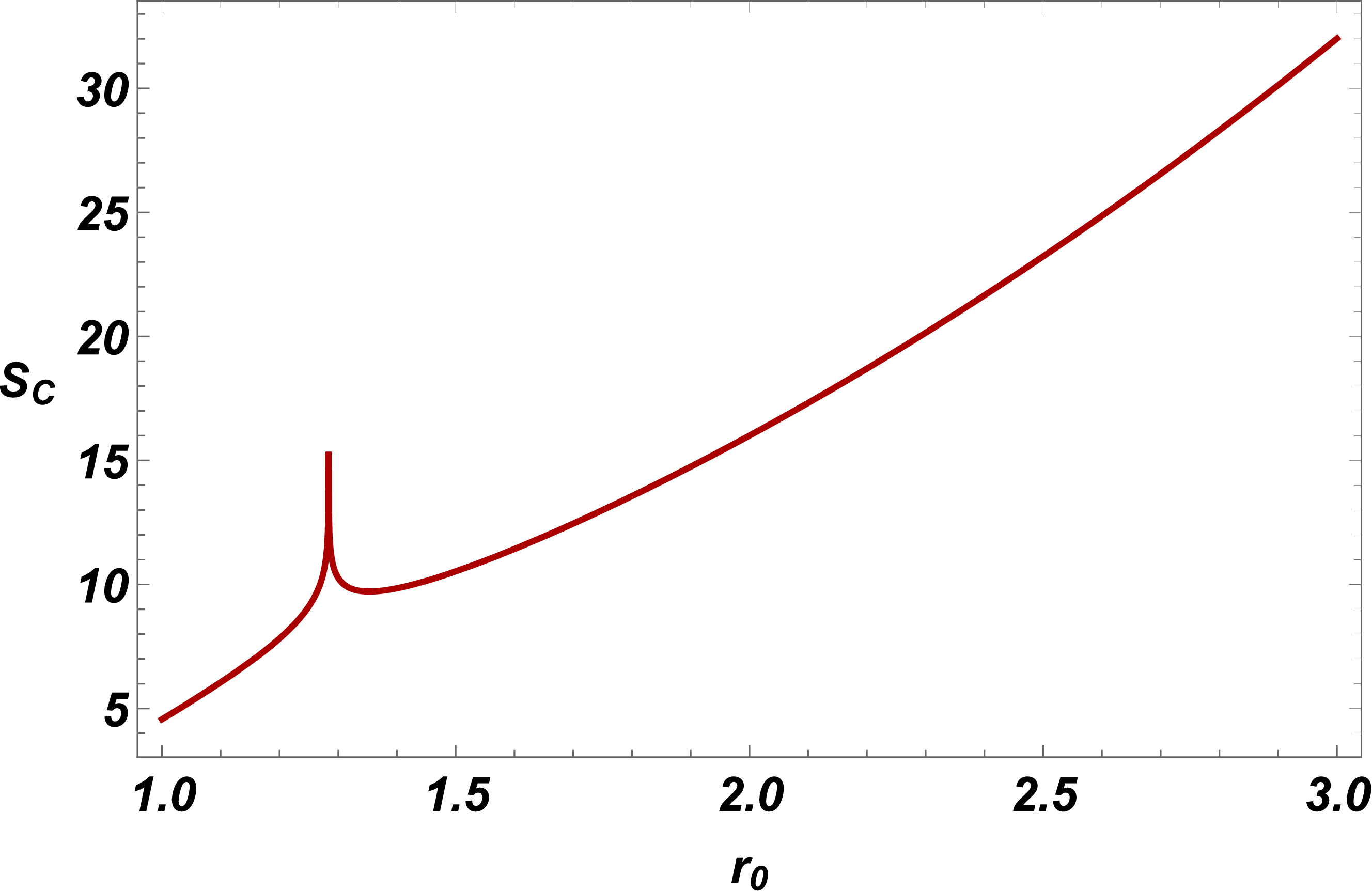}\\
    \includegraphics[scale=0.12]{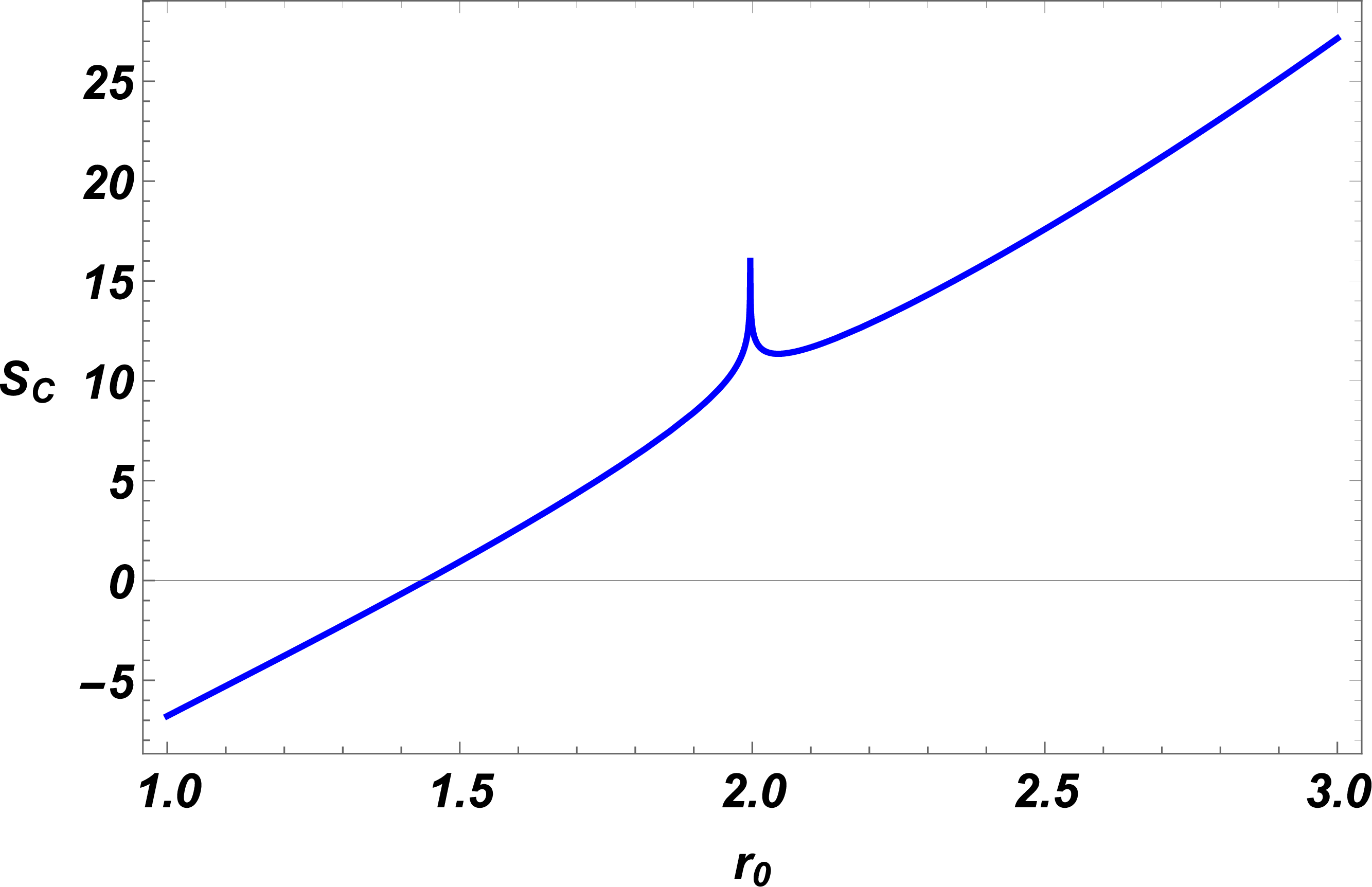}
     \includegraphics[scale=0.12]{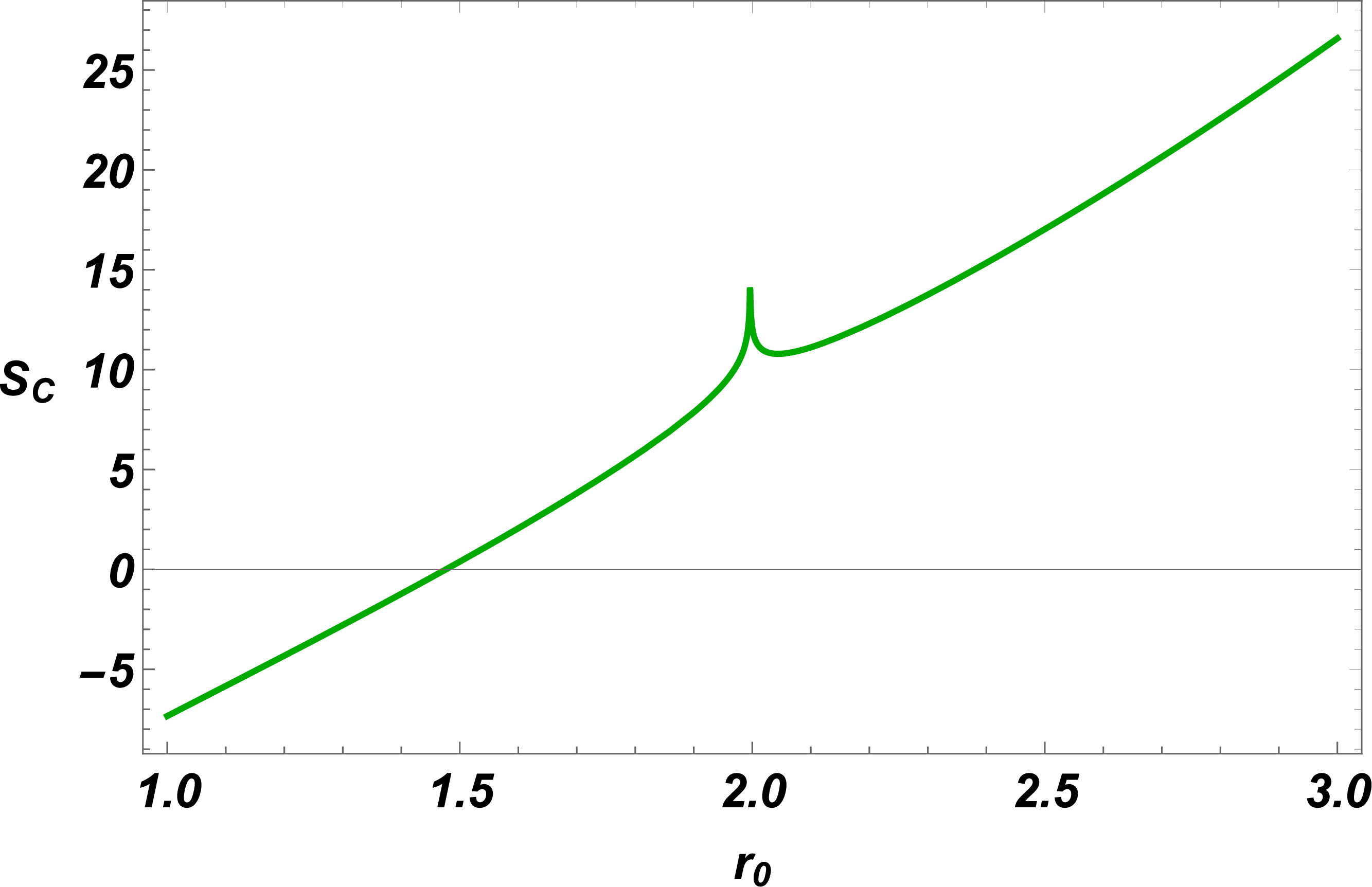}
    \includegraphics[scale=0.12]{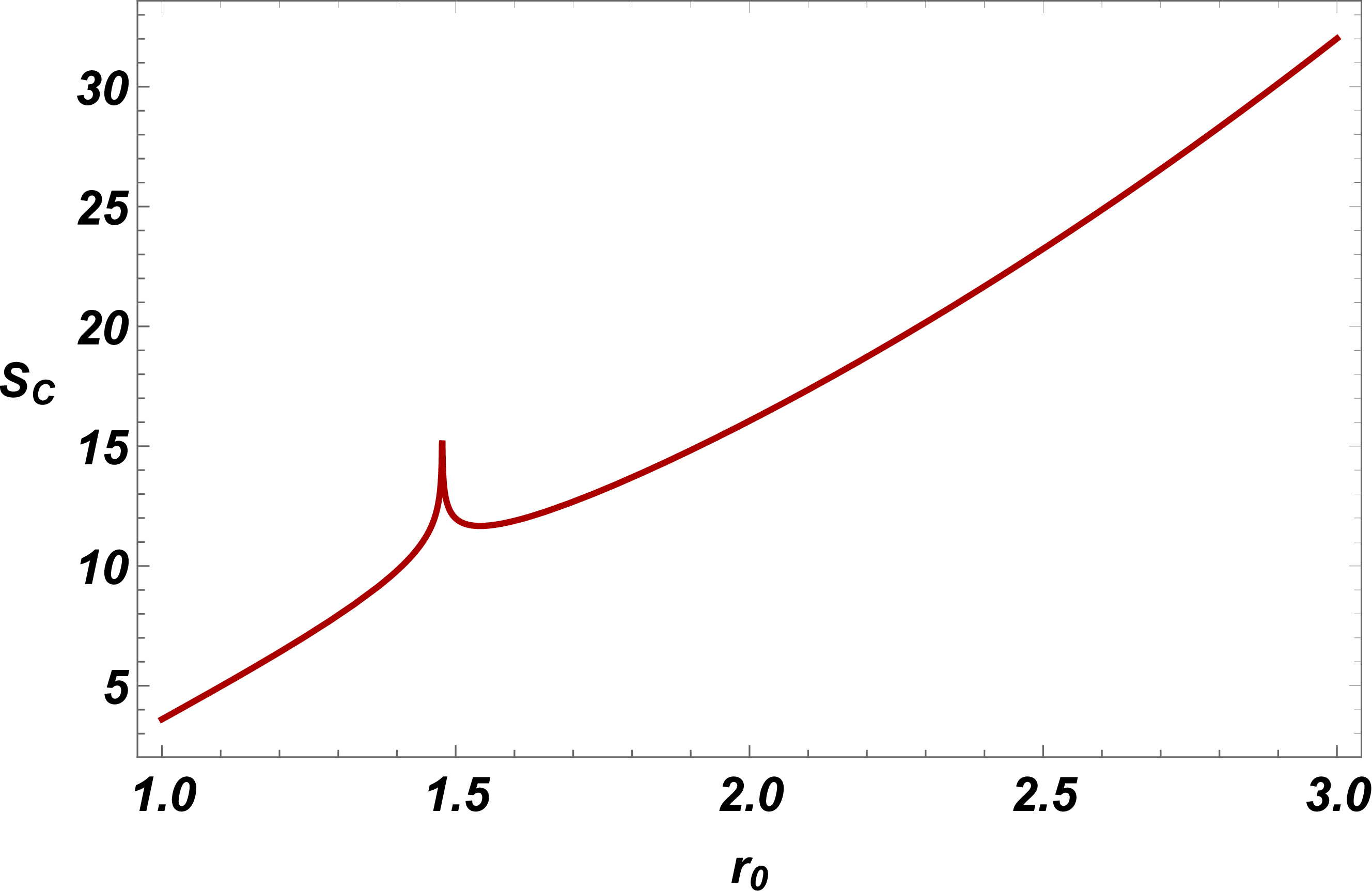}
    \caption{corrected entropy $S_c$ along horizon radius $r_0$. Herein we have fixed values $M=1,\;G=1\;\gamma=0.4$.\{Upper panel $\gamma=0$ (Left graph), $\gamma=0.2$ (Middle graph), $\gamma=0.4$ (Right graph)\} and \{Lower panel $\beta=0.2$ (Left graph), $\beta=0.4$ (Middle graph), $\beta=0.6$ (Right graph)\}}
    \label{plot:5}
    \end{figure}

In this section, we investigate the impact of thermal fluctuations on the thermodynamics of the static BH in the SBRG. To study this phenomenon, we utilize the formalism of Euclidean quantum gravity, which involves rotating the temporal coordinate within a complex plane. As a result, the partition function for a BH can be mathematically written as follows \cite{ad24,ad25,ad26,ad27}:
\begin{equation}\label{1b}
    Z=\int D g D A \exp (-I)
\end{equation}
where $I \rightarrow \iota I$ is Euclidean action for this system. One can relate the statistical mechanical partition function \cite{ad28,ad29} as
\begin{equation}\label{2b}
    Z=\int_0^{\infty} D E \Gamma(E) \exp (-\psi E)
\end{equation}
where $\psi=T^{-1}$. We can calculate the density of states by using
\begin{equation}\label{3b}
    \Gamma(E)=\frac{1}{2 \pi \iota} \int_{\psi_0-\iota \infty}^{\psi_0+ \iota \infty} d \psi e^{S(\psi)}
\end{equation}
where $S_c=\psi E+\ln Z$. The entropy near the equilibrium temperature $\psi$ can be calculated by neglecting thermal fluctuations, which yields $S = \pi r_{0}^2$. But, when we are taking into account thermal fluctuations then the entropy $S_c(\psi)$ is given by: \cite{ad24}
\begin{equation}\label{4b}
    S_c=S+\frac{1}{2}\left(\psi-\psi_0\right)\left(\frac{\partial^2 S(\psi)}{\partial \psi^2}\right)_{\psi=\psi_0}
\end{equation}
So, one can write the density as stated by:
\begin{equation}\label{5b}
    \Gamma(E)=\frac{1}{2 \pi \iota} \int_{\psi_0-\iota \infty}^{\psi_0+\iota \infty} d \psi e^{\frac{1}{2}\left(\psi-\psi_0\right)\left(\frac{\partial^2 S(\psi)}{\partial \psi^2}\right)_{\psi=\psi_0},}
\end{equation}
which leads to
\begin{equation}\label{6b}
    \Gamma(E)=\frac{e^{S}}{\sqrt{2 \pi}}\left[\left(\frac{\partial^2 S(\psi)}{\partial \psi^2}\right)_{\psi=\psi_0}\right]^{\frac{1}{2}}
\end{equation}
We can write corrected entropy as:
\begin{equation}\label{7b}
    S_c=S-\frac{1}{2} \ln \left[\left(\frac{\partial^2 S(\psi)}{\partial \psi^2}\right)_{\psi=\psi_0}\right]^{\frac{1}{2}}.
\end{equation}
The second derivative of entropy measures the squared fluctuation of energy. By utilizing the connection between conformal field theory and the microscopic degrees of freedom of a BH \cite{ad30} it becomes possible to simplify this expression. Consequently, the entropy can be expressed as $S = m_1 \psi^{n_1} + m_2 \psi^{-n_2}$, where $m_1$, $m_2$, $n_1$, and $n_2$ are positive constants \cite{ad31}. This entropy exhibits an extremum at $\psi_0 = \left(\frac{m n_2}{m_1 n_1}\right)^{\frac{1}{n_1+n_2}} = T^{-1}$, where $T$ denotes temperature. By expanding the entropy around this extremum, we can determine \cite{ad32,ad33,f3}:
\begin{equation}\label{8b}
    \left(\frac{\partial^2 S(\psi)}{\partial \psi^2}\right)_{\psi=\psi_0}=S \psi_0^{-2} .
\end{equation}
A corrected version of entropy by negating higher-order correction terms can be written as:
\begin{equation}\label{9b}
    S_c=S-\frac{1}{2} \ln S T^2
\end{equation}
Furthermore, the presence of quantum fluctuations in the geometry of BHs introduces a notable concern regarding thermal fluctuations in BH thermodynamics. These correction terms become significant when the size of the BH is small and its temperature is large. Hence, for large BHs, quantum fluctuations can be disregarded. It becomes evident that thermal fluctuations become significant solely for BHs characterized by high temperatures, and as the BH size decreases, its temperature increases. Thus, we can deduce that these correction terms are applicable solely to sufficiently small BHs exhibiting high temperatures \cite{ad24}.
Subsequently, we can derive the general expression for the entropy by neglecting higher-order correction terms:
\begin{equation}\label{10b}
    S_c=S-\gamma \ln S T^2.
\end{equation}
Here, we introduce $\gamma$ as a constant parameter to incorporate the logarithmic correction terms associated with thermal fluctuations. By setting $\gamma=0$, we recover the entropy without any correction terms. As mentioned earlier, in the case of large BHs with extremely low temperatures, we can take $\gamma \rightarrow 0$, whereas, for small BHs with sufficiently high temperatures, we consider $\gamma \rightarrow 1$. By utilizing Eqs. (\ref{9aa}) and (\ref{10b}), we can obtain the following expression for the corrected entropy:
\begin{equation}
    S_c=\pi  r_{0}^{2}-\gamma  \log \left(\frac{\left(993280 \pi ^3 \sqrt{2} \beta  G^7 M^4-497664 \sqrt{2} \pi ^3 \beta  G^6 M^3 r_{0}+5 G M r_{0}^{9}\right)^2}{100 \pi  r_{0}^{20}}\right).
\end{equation}
 We plot $S_C$ considering different choices of the correction parameter: $\gamma=0$ for large BHs, $\gamma=0.2,\;0.4\;$ and  $\;0.6$ for small BHs ($0<\gamma<1$). The plot shown in Fig. (\ref{plot:5}) demonstrates that the entropy of spacetime consistently increases across the entire range considered for different values of $\gamma$ and the parameter $\beta$. Notably, these fluctuations are more pronounced for smaller BHs. 

\section{Lensing with plasma}

Weak gravitational lensing has proven to be a noteworthy optical phenomenon in the investigation of BHs. In this context, our current focus is on exploring the weak gravitational gravitational lensing within the framework of the plasma field for a static BH in the SBRG. The metric tensor that governs the weak field approximation can be expressed as follows \cite{ad34}
\begin{equation}
    g_{\alpha \lambda}=\eta_{\alpha \lambda}+h_{\alpha \lambda}\,
\end{equation}
where $\eta_{\alpha \lambda}$ and $h_{\alpha \lambda}$ represent the Minkowski spacetime and perturbation gravity field, respectively. These terms must satisfy the following properties:
\begin{eqnarray}
 &&   \eta_{\alpha \lambda}=diag(-1,1,1,1)\ , \nonumber\\
 &&   h_{\alpha \lambda} \ll 1, \hspace{0.5cm} h_{\alpha \lambda} \rightarrow 0 \hspace{0.5cm} under\hspace{0.2cm}  x^{\alpha}\rightarrow \infty \ ,\nonumber\\
 &&     g^{\alpha \lambda}=\eta^{\alpha \lambda}-h^{\alpha \lambda}, \hspace{0,5cm} h^{\alpha \lambda}=h_{\alpha \lambda}.
\end{eqnarray}
The angle of deflection around the BH can be obtained by varying the above basic equations given by
\begin{equation}
    \hat{\alpha }_{\text{b}}=\frac{1}{2}\int_{-\infty}^{\infty}\frac{b}{r}\left(\frac{dh_{33}}{dr}+\frac{1}{1-\omega^2_e/ \omega}\frac{dh_{00}}{dr}-\frac{K_e}{\omega^2-\omega^2_e}\frac{dN}{dr} \right)dz\, 
\end{equation}
where $\omega$ and $\omega_{e}$ subsequently stand for the photon and plasma frequencies. The components $h_{\alpha \lambda}$ in the form of Cartesian coordinates using lapse function Eq. \ref{5} yield as: 
\begin{eqnarray}
     h_{00}&=\left(\frac{12416 \pi ^3 \sqrt{2} \beta  G^3 R_{s}^4}{5 r^{10}}-\frac{13824 \sqrt{2} \pi ^3 \beta  G^3 R_{s}^3}{5 r^9}+\frac{R_{s}}{r}\right) ,\label{h3}\\
    h_{ik}&=\left(\frac{12416 \pi ^3 \sqrt{2} \beta  G^3 R_{s}^4}{5 r^{10}}-\frac{13824 \sqrt{2} \pi ^3 \beta  G^3 R_{s}^3}{5 r^9}+\frac{R_{s}}{r}\right)n_i n_k,\label{h2} \\
    h_{33}&=\left(\frac{12416 \pi ^3 \sqrt{2} \beta  G^3 R_{s}^4}{5 r^{10}}-\frac{13824 \sqrt{2} \pi ^3 \beta  G^3 R_{s}^3}{5 r^9}+\frac{R_{s}}{r}\right)\cos^2\chi \label{h}.\label{h3}
    \end{eqnarray}
The deflection angle can be expressed by the following relation \cite{ad35}
\begin{equation}
    \hat{\alpha}_{b}=\hat{\alpha}_{1}+\hat{\alpha}_{2}+\hat{\alpha}_{3}\ , \label{h6}
\end{equation}
where
 \begin{eqnarray}
\hat{\alpha}_{1}&=&\frac{1}{2}\int_{-\infty}^{\infty} \frac{b}{r}\frac{dh_{33}}{dr}dz\ ,\nonumber\\
\hat{\alpha}_{2}&=&\frac{1}{2}\int_{-\infty}^{\infty} \frac{b}{r}\frac{1}{1-\omega^2_e/ \omega}\frac{dh_{00}}{dr}dz\ ,\nonumber\\
\hat{\alpha}_{3}&=&\frac{1}{2}\int_{-\infty}^{\infty} \frac{b}{r}\left(-\frac{K_e}{\omega^2-\omega^2_e}\frac{dN}{dr} \right)dz\ .  \label{h7}
\end{eqnarray}

Furthermore, we compute the deflection angle for both uniform and non-uniform plasma density distributions.

It may be worth noting that in our further discussions, we use $\omega$ instead of $\omega(\infty)$ and $\omega_0$ instead of $\omega_e(\infty)$ \cite{ad34,ad36}.

\subsection{Uniform plasma}\label{A8.1}
The deflection angle for uniform plasma around the BH turns out to be \cite{ad35} 
\begin{equation}\label{h8}
  \hat{\alpha}_{uni}=\hat{\alpha}_{uni1}+\hat{\alpha}_{uni2}+\hat{\alpha}_{uni3}.
\end{equation}
Solving Eqs.(\ref{h3}), (\ref{h6}) and (\ref{h7}), we find the deflection angle in the uniform plasma given by

 \begin{figure}
    \includegraphics[scale=0.5]{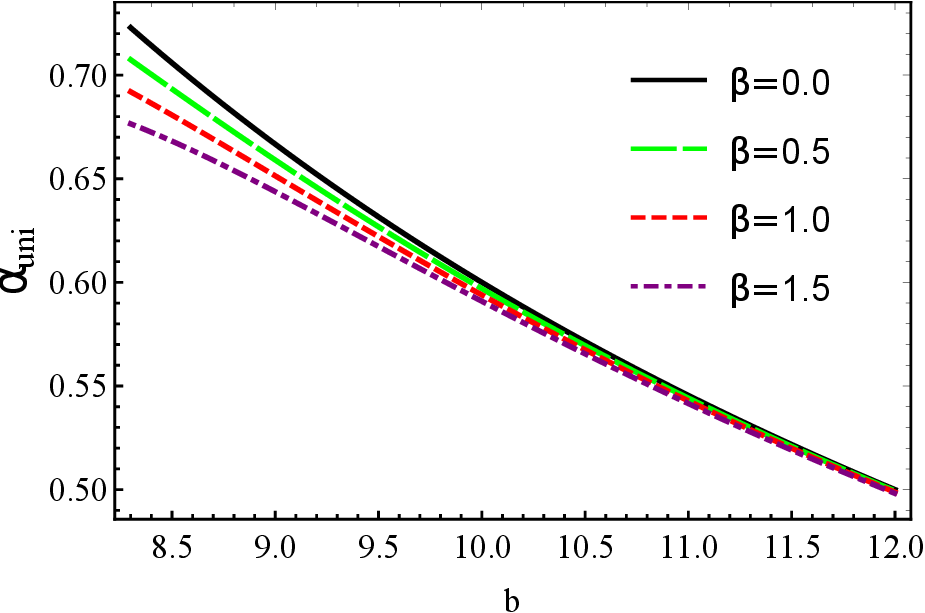}
    \includegraphics[scale=0.5]{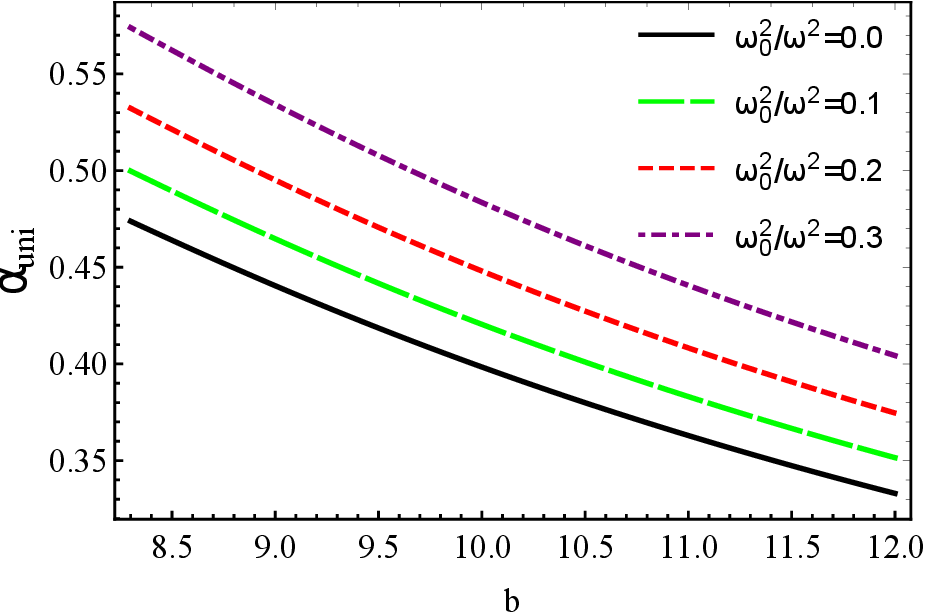}
    \caption{Deflection angle for uniform plasma with fixed values $G=1,\;\beta=0.5,\;R_s=2,\;\omega_02/\omega^2=0.5,\;b=7.0.$}
    \label{plot:7}
    \end{figure}

     \begin{figure}
    \includegraphics[scale=0.5]{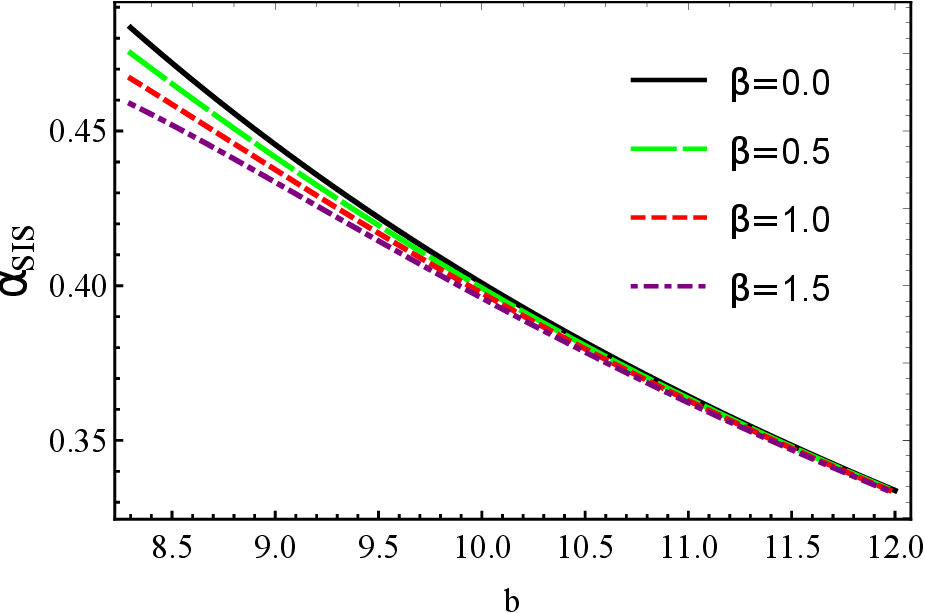}
    \includegraphics[scale=0.5]{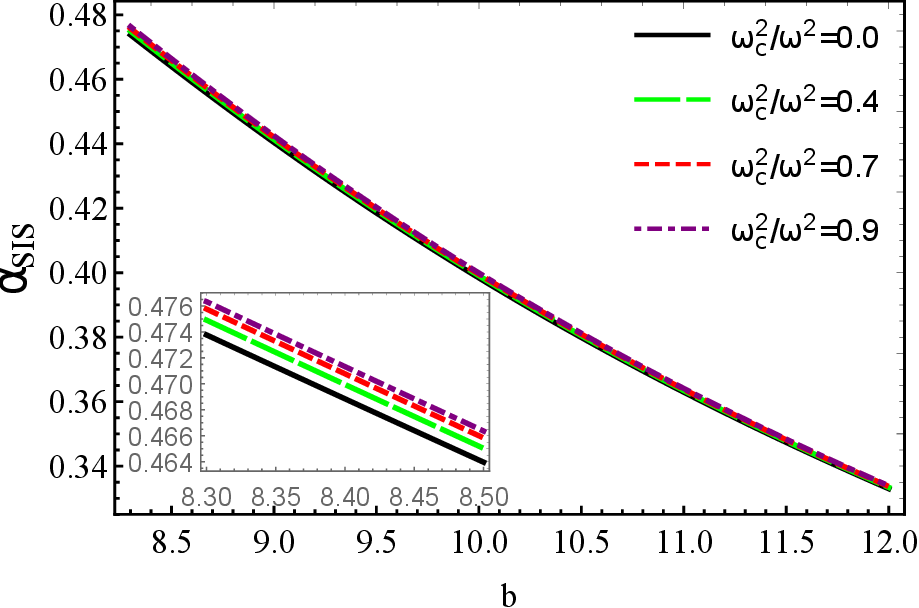}
    \caption{Deflection angle for $SIS$ plasma with fixed values $G=1,\;\beta=0.5,\;R_s=2,\;\omega_02/\omega^2=0.5,\;b=7.0.$}
    \label{plot:8}
    \end{figure}
      \begin{figure}
   \includegraphics[scale=0.18]{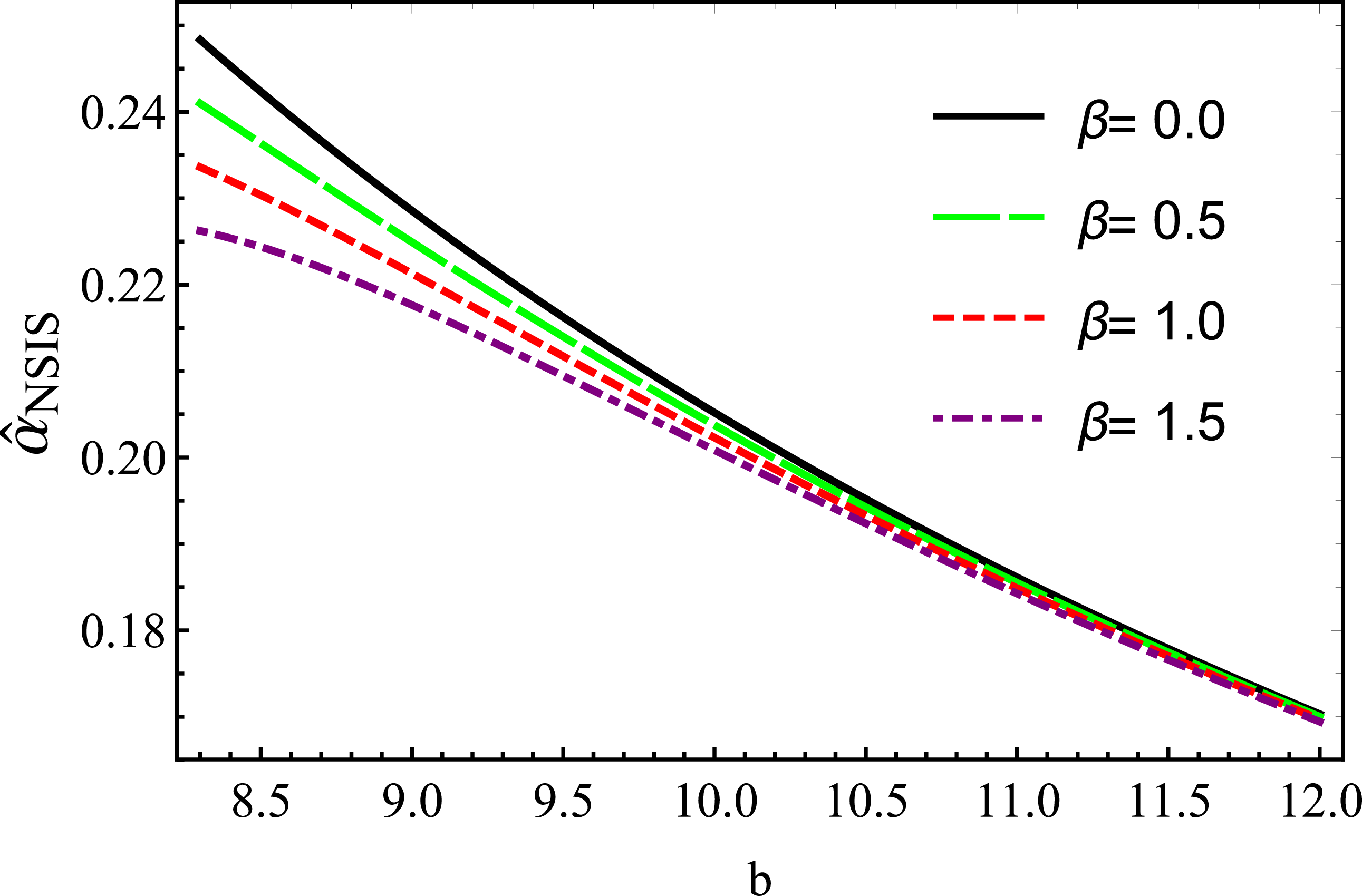}
    \includegraphics[scale=0.18]{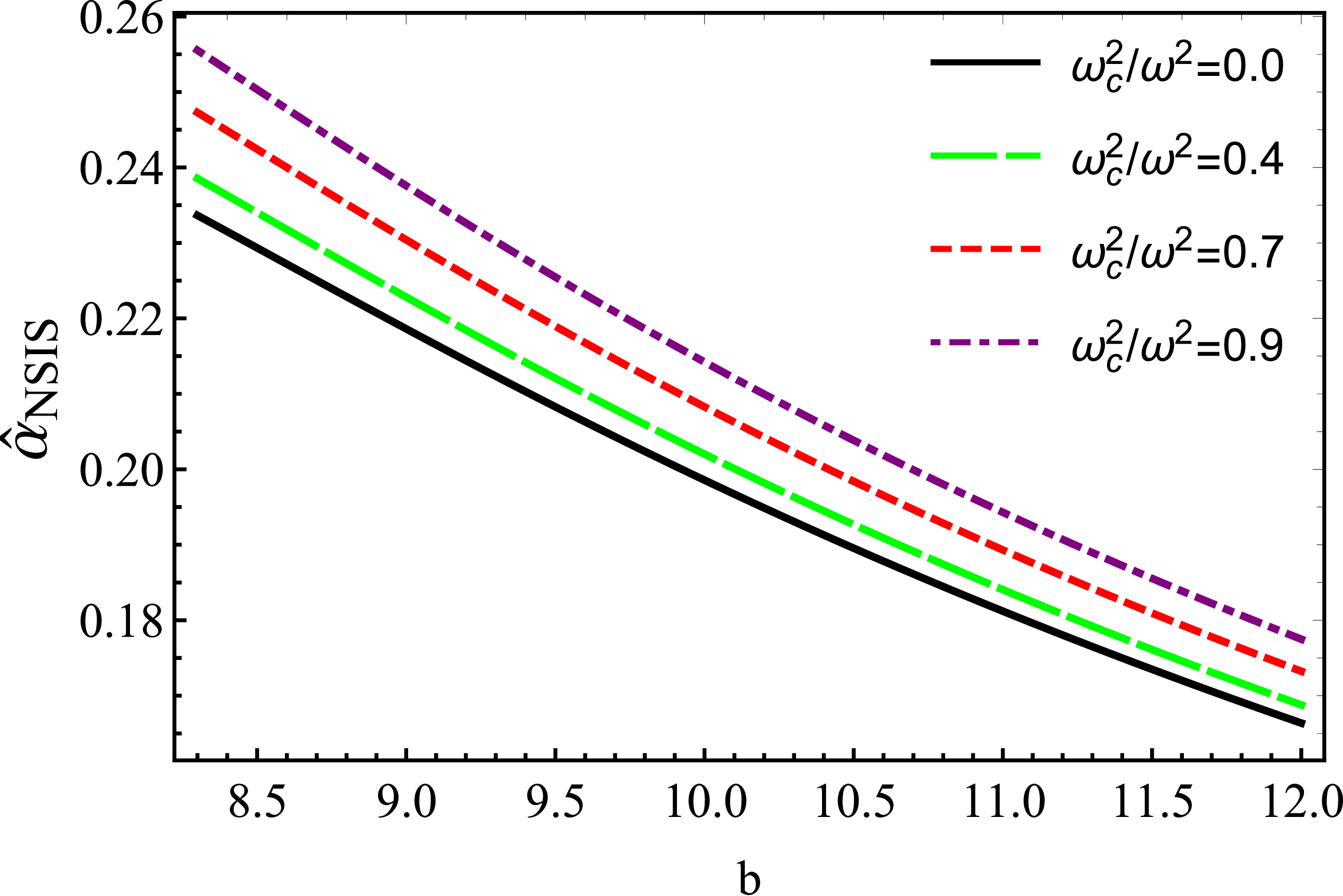}
    \caption{Deflection angle for $NSIS$ plasma with fixed values $G=1,\;\beta=0.5,\;R_s=2,\;\omega_{c}^2/\omega^2=0.5,\;b=7.0.$}
    \label{plot:9a}
    \end{figure}

  \begin{figure}
    \includegraphics[scale=0.12]{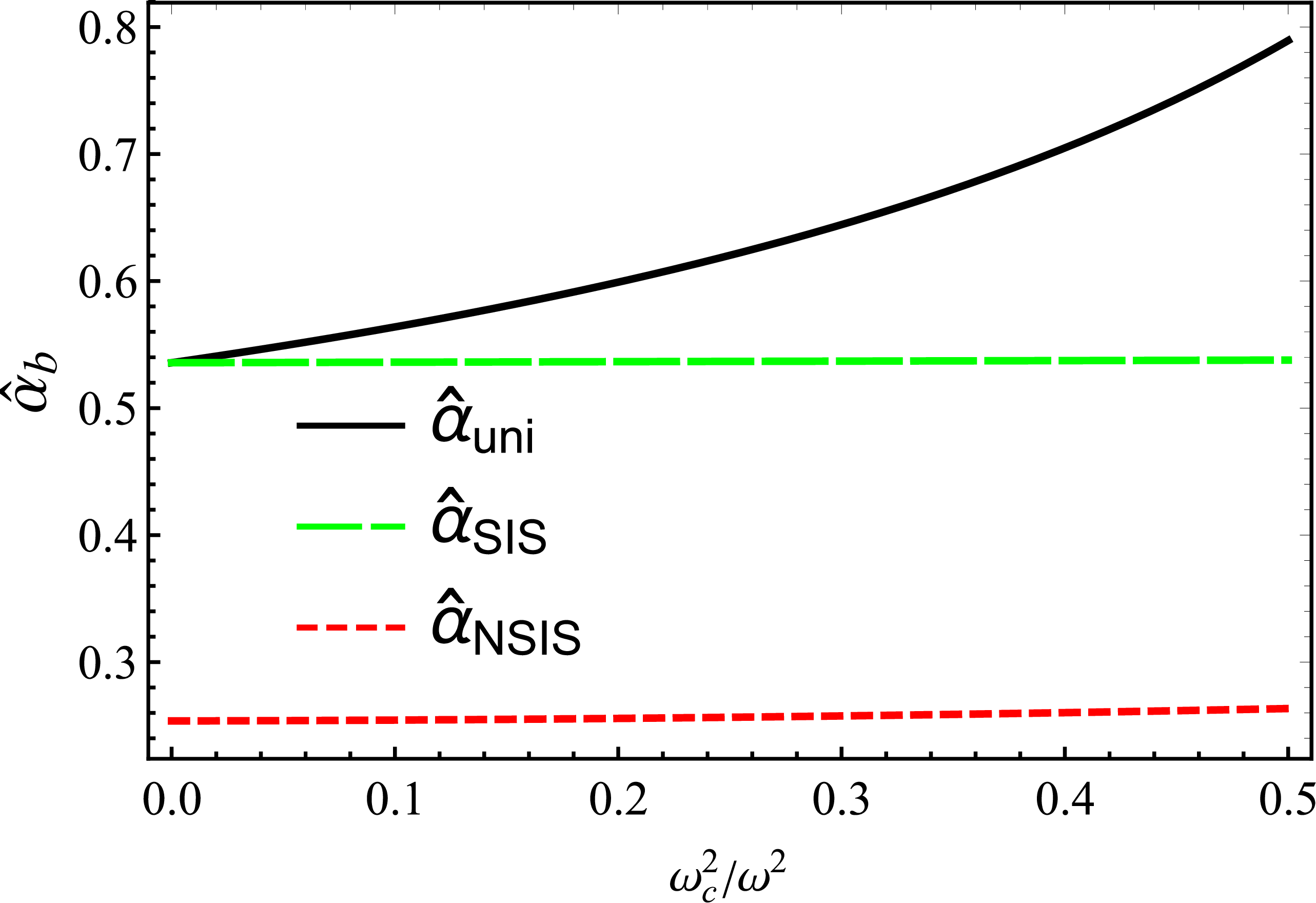}
    \includegraphics[scale=0.12]{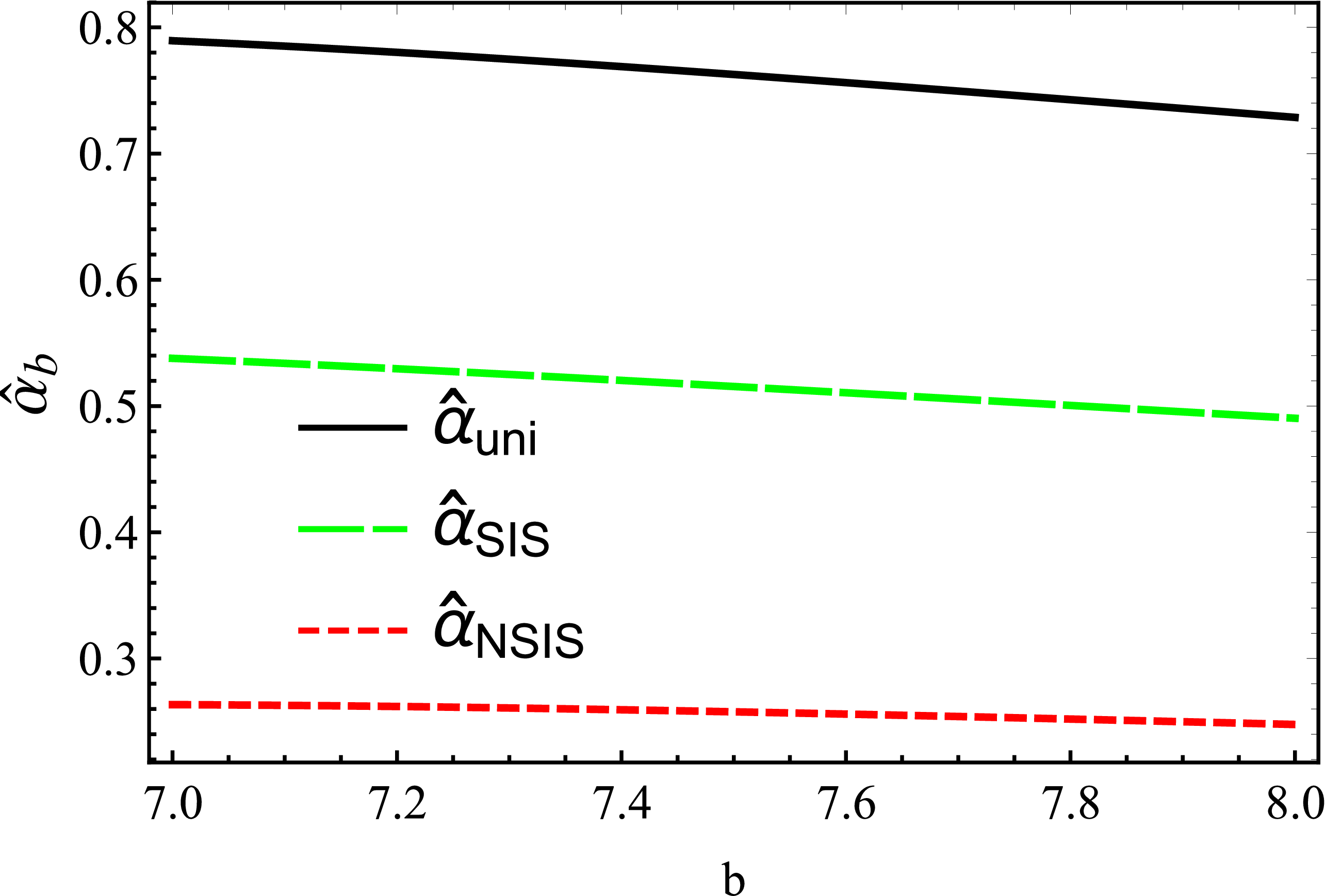}
     \includegraphics[scale=0.12]{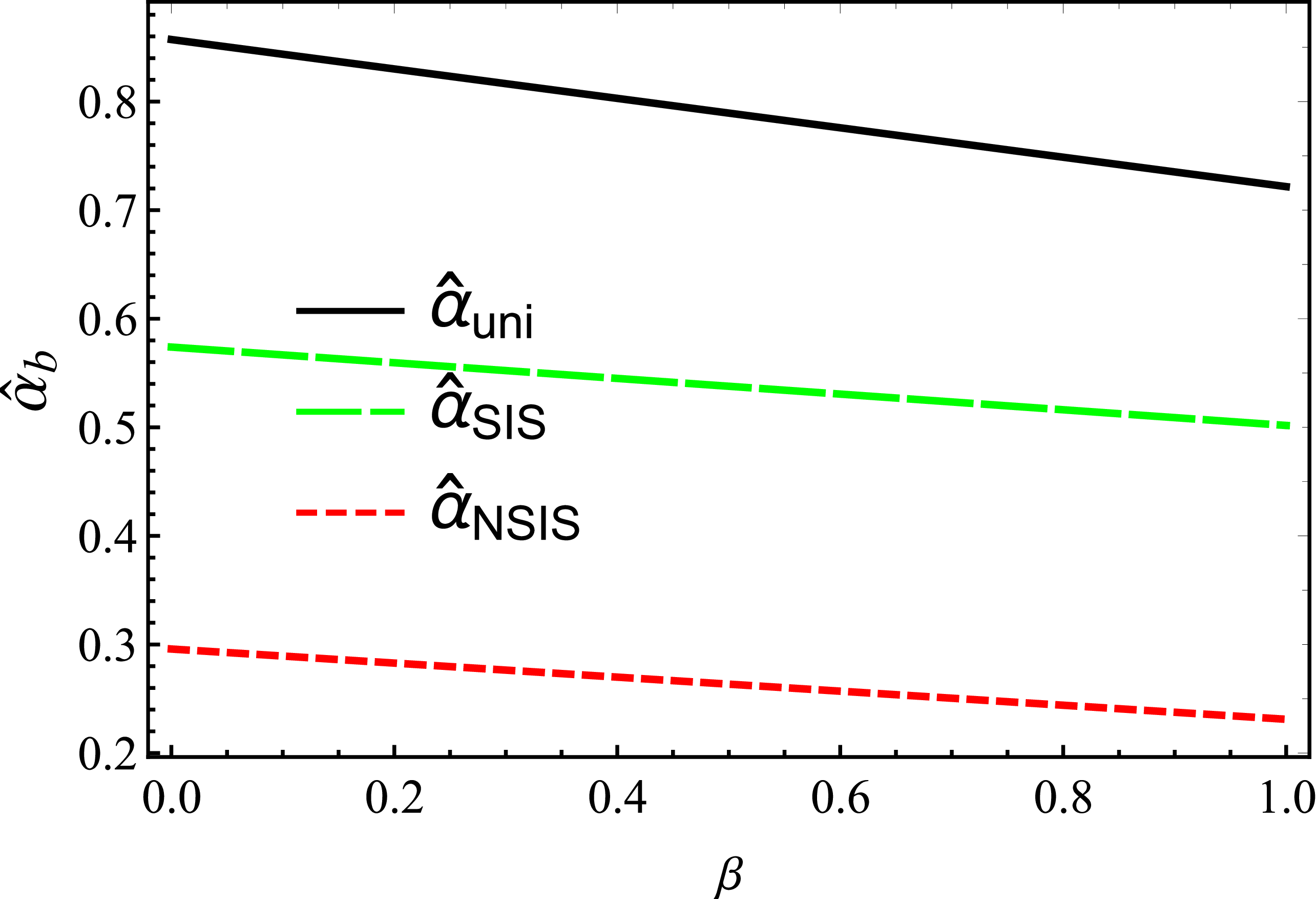}
    \caption{Deflection angle for uniform, $SIS$ and $NSIS$ plasma with fixed values $G=1,\;R_s=2,\;\beta=0.5,\;\omega_{0}^2/\omega^2=0.5,\;b=7.0.$}
    \label{plot:10}
    \end{figure}
     \begin{figure}
    \includegraphics[scale=0.18]{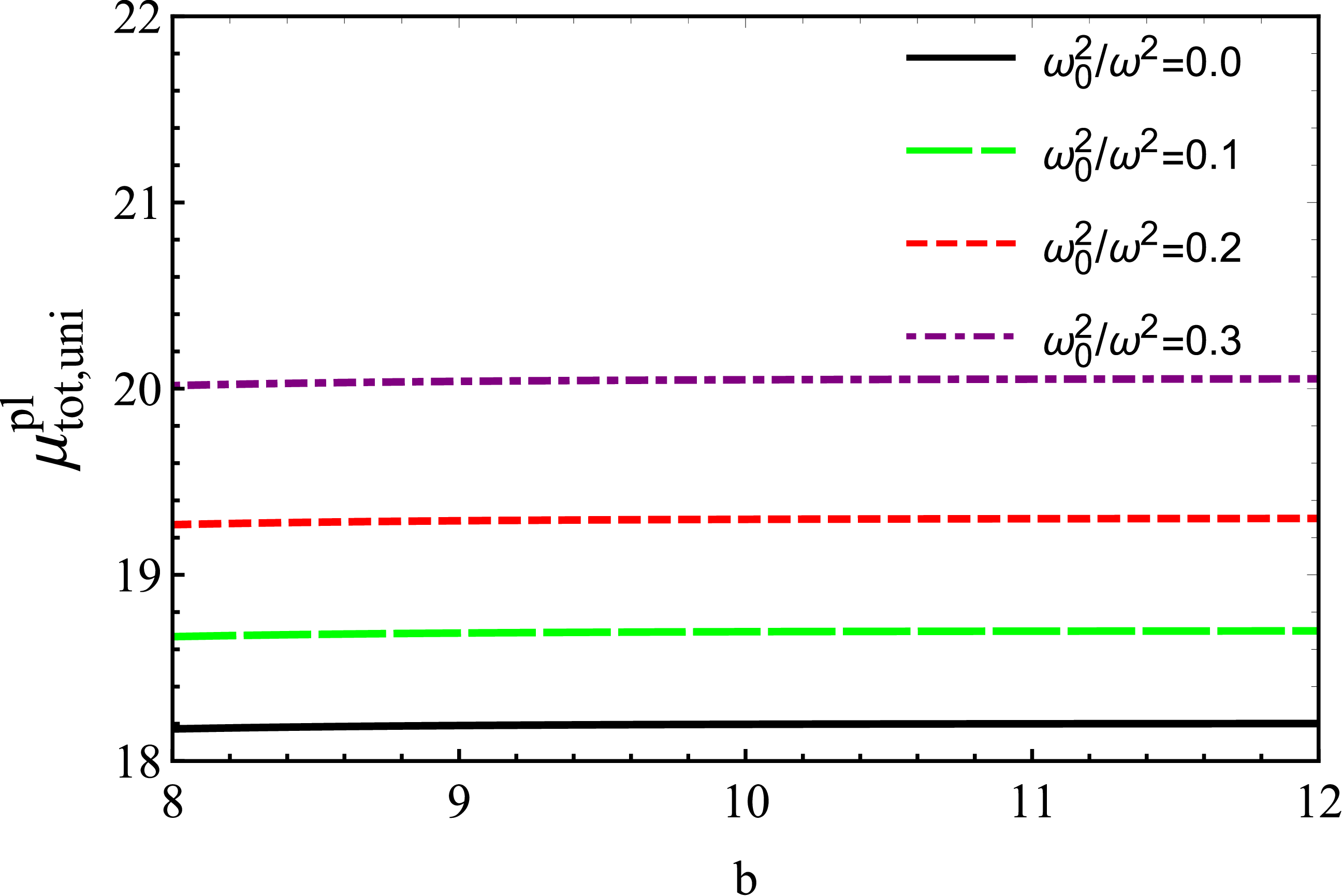}
    \includegraphics[scale=0.18]{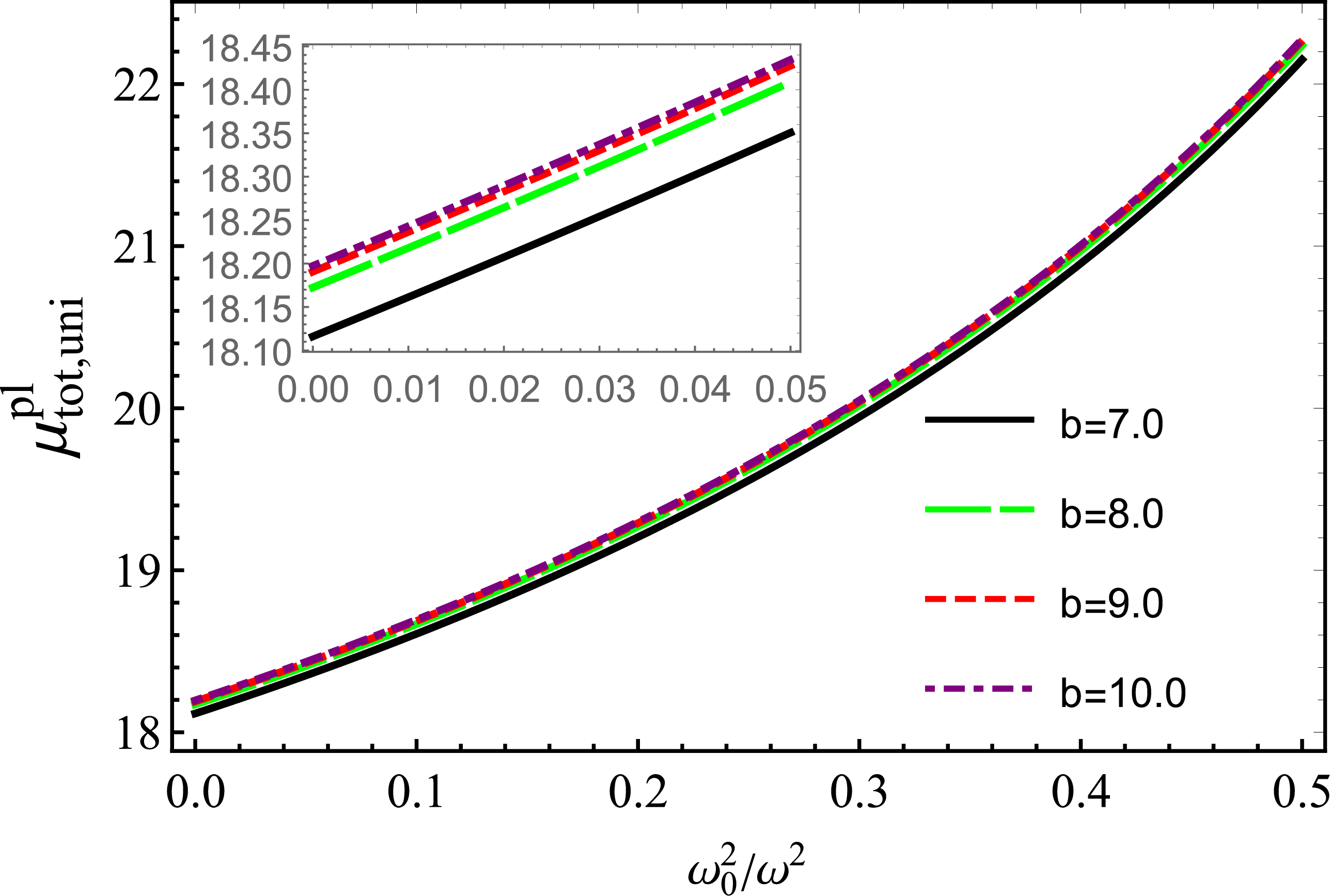}
     \includegraphics[scale=0.18]{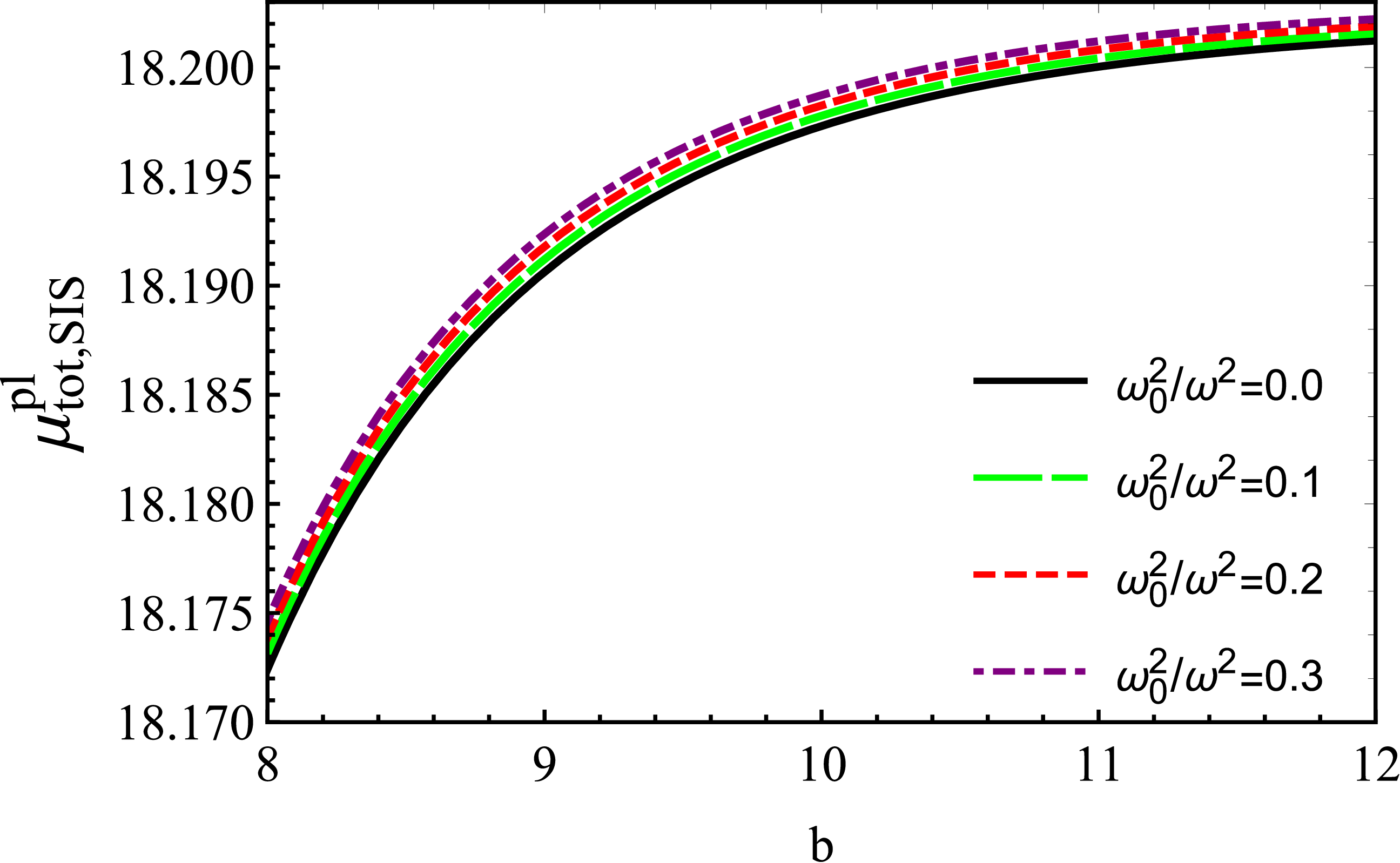}
     \includegraphics[scale=0.18]{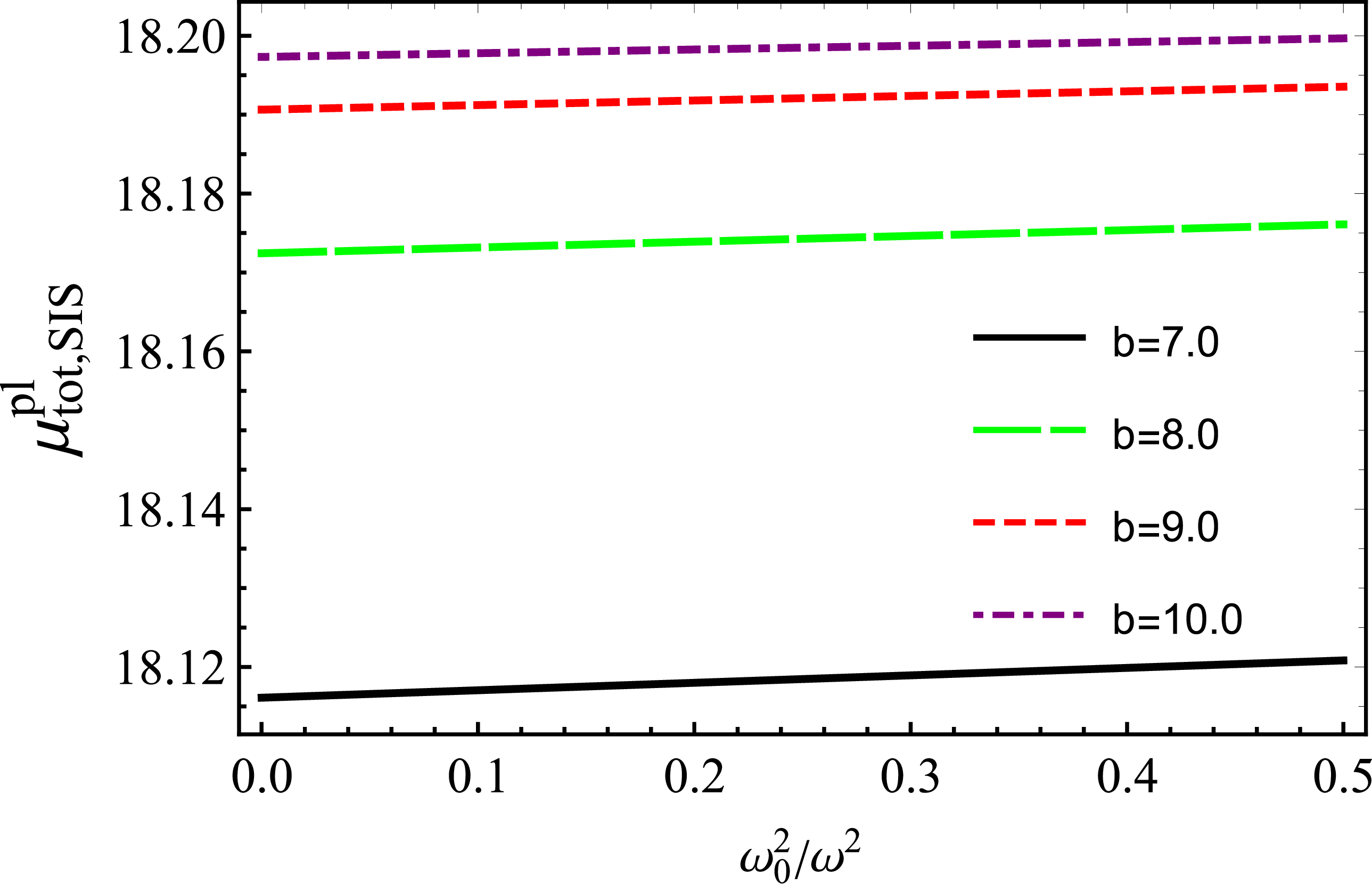}
    \caption{Total magnification for uniform (Upper graphs) and $SIS$ (Lower graphs) plasma with fixed values $G=1,\;\beta=0.5,\;R_s=2,\;\omega_02/\omega^2=0.5,\;b=7.0.$}
    \label{plot:11}
    \end{figure}
     \begin{figure}
    \includegraphics[scale=0.12]{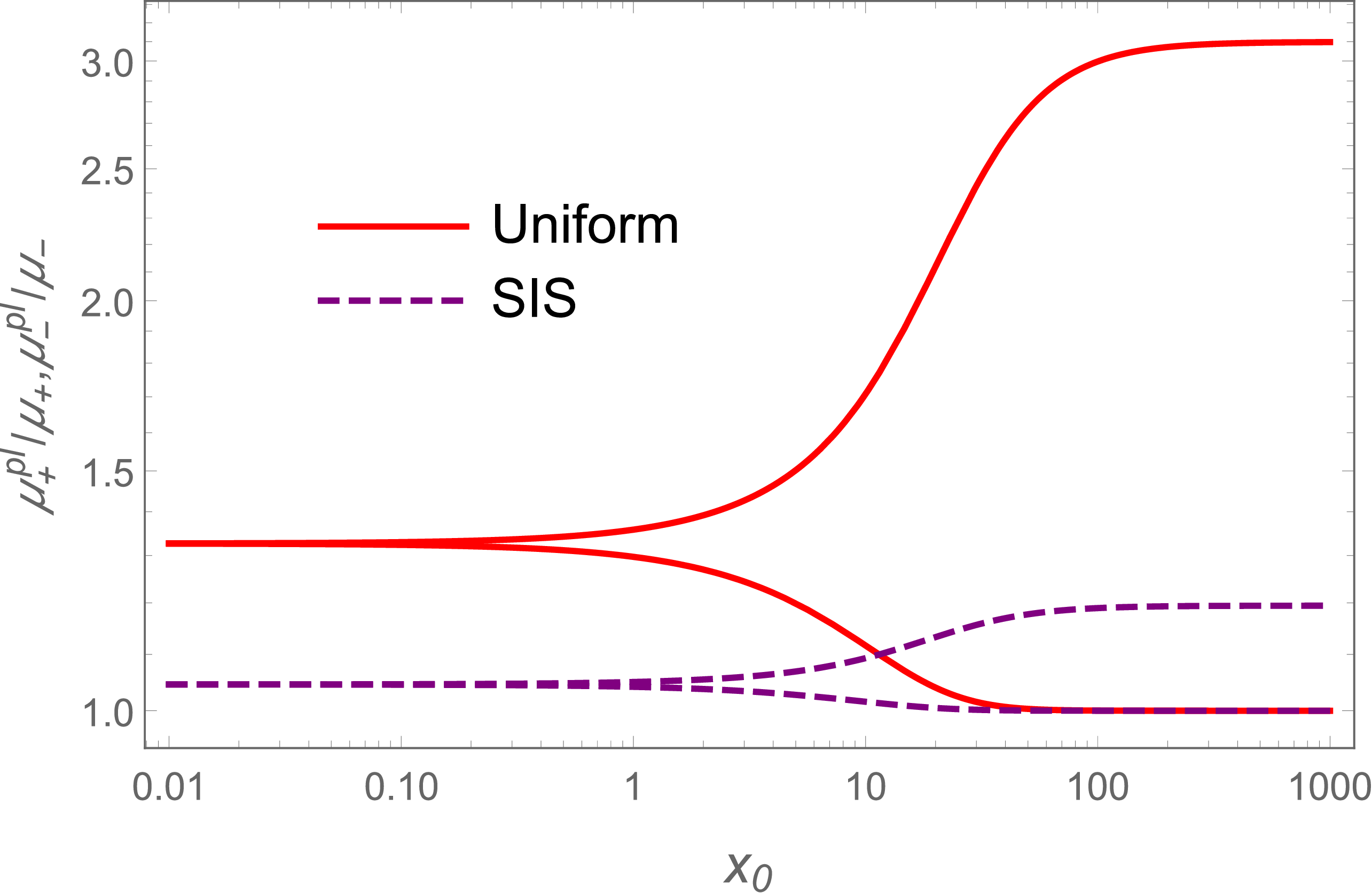}
    \includegraphics[scale=0.12]{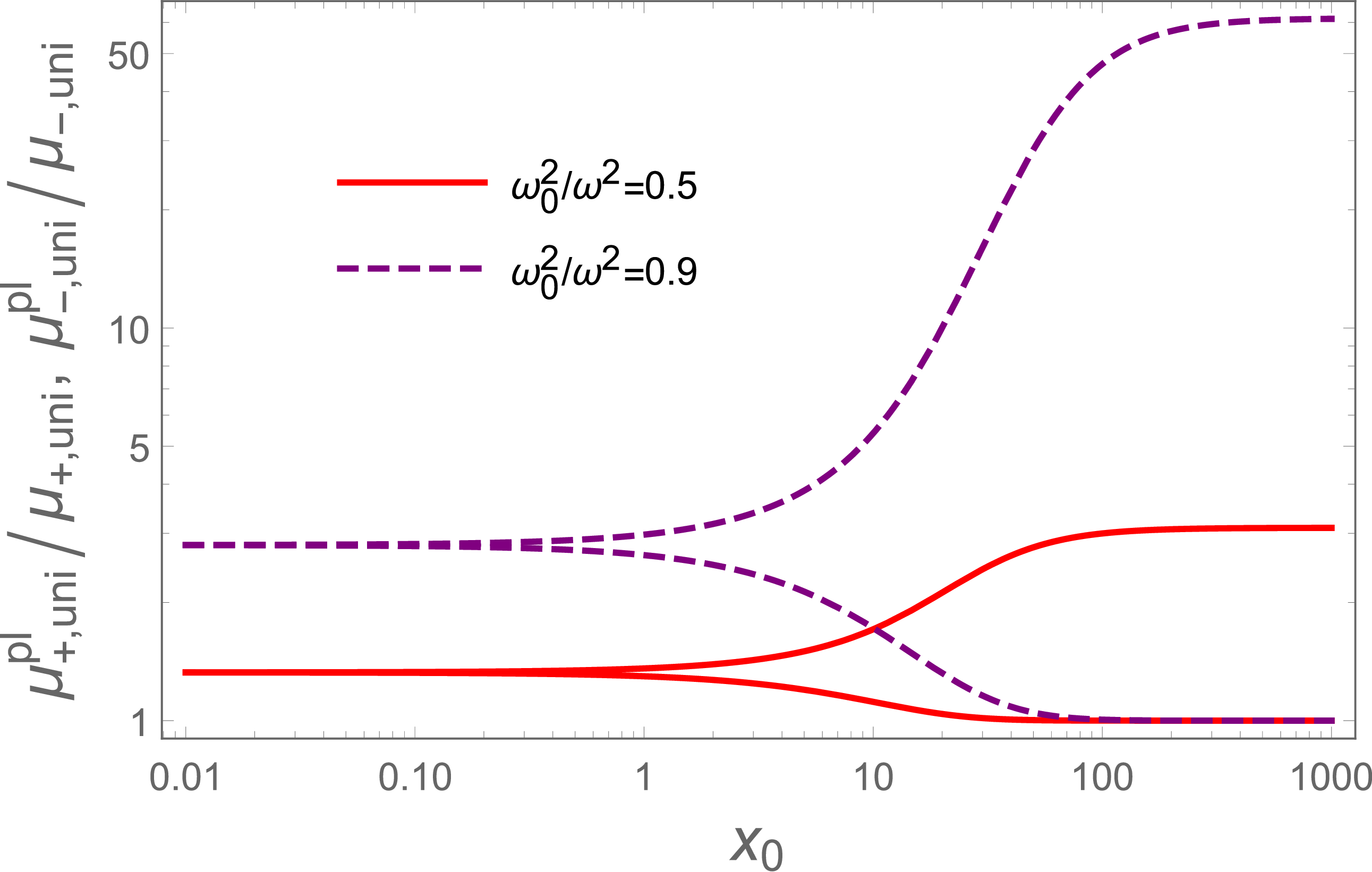}
     \includegraphics[scale=0.12]{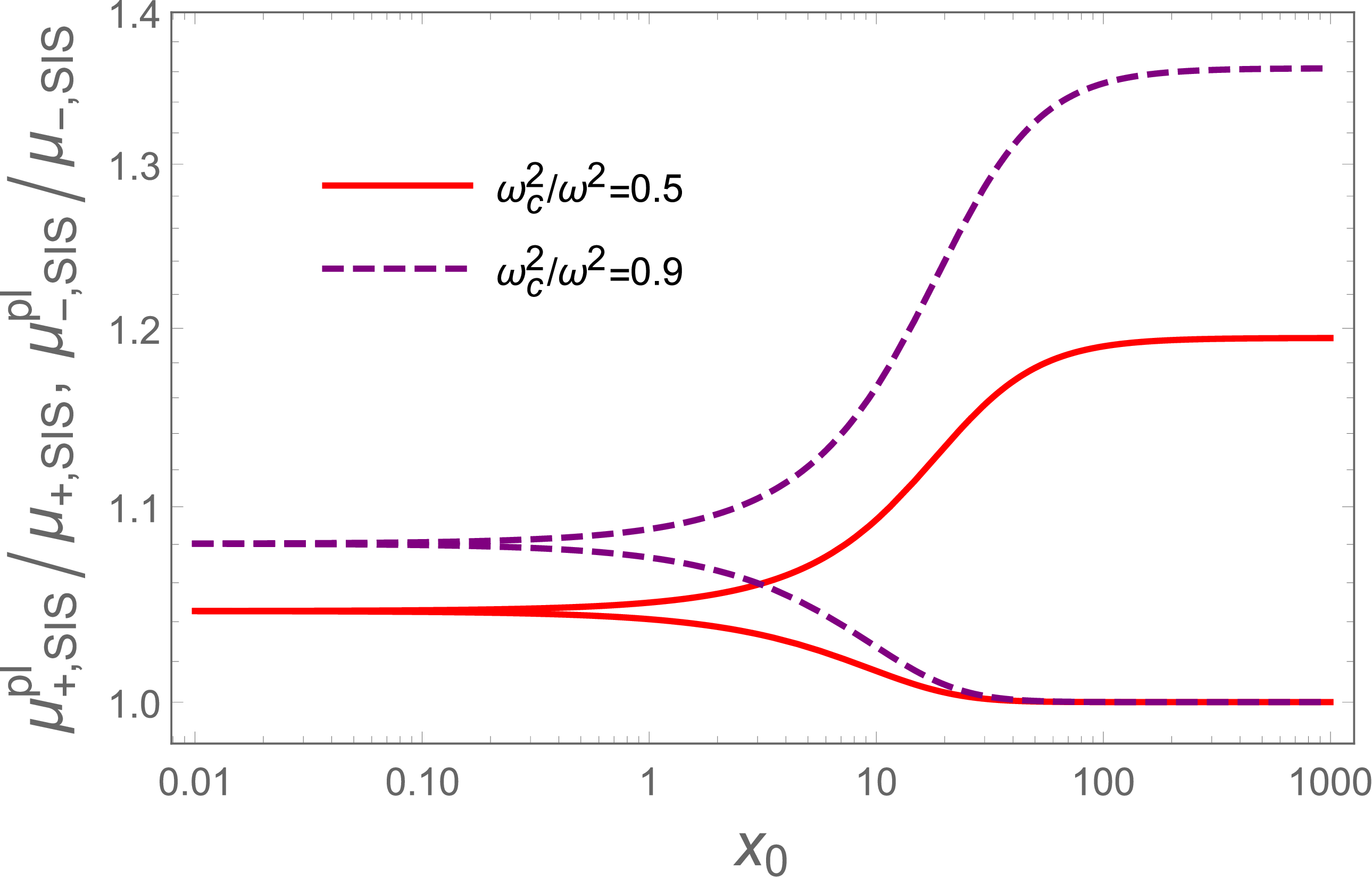}
    \caption{Deflection angle for uniform, $SIS$ and $NSIS$ plasma with fixed values $G=1,\;\beta=0.5,\;R_s=2,\;\omega_02/\omega^2=0.5,\;b=7.0.$}
    \label{plot:12}
    \end{figure}
 \begin{eqnarray}\label{40}
    \hat{\alpha}_{uni}=\frac{196608 \pi ^3 \sqrt{2} \beta  G^3 R_{s}^3}{175 b^9}-\frac{6111 \pi ^4 \sqrt{\frac{1}{b^2}} \beta  G^3 R_{s}^4}{10 \sqrt{2} b^9}+\frac{\frac{1769472 \pi ^3 \sqrt{2} \beta  G^3 R_{s}^3}{175 b^9}-\frac{6111 \pi ^4 \sqrt{\frac{1}{b^2}} \beta  G^3 R_{s}^4}{\sqrt{2} b^9}-\frac{R_{s}}{b}}{1-\frac{\omega_0^2}{\omega^2}}-\frac{R_{s}}{b}. 
 \end{eqnarray}
We also provide a graphical illustration of the deflection angle in Figs.~\ref{plot:7}. It is worthwhile to mention that  $\hat{\alpha}_{uni}$, along the impact parameter $b$, decreases by increasing $\beta$ and increases with an increase in $\omega_0^2/\omega^2$. The $\hat{\alpha}_{uni}$ along $\beta$ decreases by increasing $b$ and increases with an increase in $\omega_0^2/\omega^2$. The $\hat{\alpha}_{uni}$ along  $\omega_0^2/\omega^2$ decreases by increasing both $\beta \; \& \;b$.
 \subsection{Non uniform plasma}\label{A8.2}
The non-uniform distribution for the plasma field is given by \cite{ad34}
\begin{equation}\label{pl}
    \rho(r)=\frac{\sigma^2_{\nu}}{2\pi r^2}\ ,
\end{equation}
where $\sigma^2_{\nu}$ defines the uni-dimensional velocity dispersion. We can write the non-uniform concentration of the plasma field as \cite{ad34}
\begin{equation}\label{pll}
    N(r)=\frac{\rho(r)}{k m_p}\ , 
\end{equation}
where $m_p$ and $k$ denote the mass and dimensionless dark matter coefficient, respectively. The plasma frequency yields
\begin{equation}
    \omega^2_e=K_e N(r)=\frac{K_e \sigma^2_{\nu}}{2\pi k m_p r^2}\ .
\end{equation}
Here we are interested to explore the influence of non-uniform plasma ($SIS$) on the deflection angle around the SBRG BH geometry. In this scenario, the deflection angle has a mathematical expression of the form \cite{ad35}
\begin{equation}
   \hat{\alpha}_{SIS}=\hat{\alpha}_{SIS1}+\hat{\alpha}_{SIS2}+\hat{\alpha}_{SIS3} \label{nonsis}\ .
\end{equation}
Using Eqs.(\ref{h3}), (\ref{h7}) and (\ref{nonsis}), one can write  the deflection angel for $SIS$ plasma as
 \begin{eqnarray}\label{40}
    \hat{\alpha}_{SIS}&=&\frac{3538944 \pi ^2 \sqrt{2} \beta  G^3 R_{s}^5 \omega_c^2}{385 b^{11} \omega^2}+\frac{393216 \pi ^3 \sqrt{2} \beta  G^3 R_{s}^3}{35 b^9}-\frac{2 R_{s}^3 \omega_c^2}{3 \pi  b^3 \omega^2}-\frac{22407 \pi ^3 \sqrt{\frac{1}{b^2}} \beta  G^3 R_{s}^6 \omega_c^2}{4 \sqrt{2} b^{11} \omega^2}\nonumber\\&-&\frac{67221 \pi ^4 \sqrt{\frac{1}{b^2}} \beta  G^3 R_{s}^4}{10 \sqrt{2} b^9}-\frac{2 R_{s}}{b}. 
 \end{eqnarray}
 The analytic form of plasma constant $\omega^2_c$ gives \cite{ad36}
 \begin{equation}
    \omega^2_c=\frac{K_e \sigma^2_{\nu}}{2\pi k m_p R^2_S}\ . 
 \end{equation}
 
 We plot the deflection angle for $SIS$ plasma in Figs.~\ref{plot:8}. Like $\hat{\alpha}_{uni}$, the $\hat{\alpha}_{SIS}$ along impact parameter $b$ decreases by increasing $\beta$ and increases with an increase in $\omega_0^2/\omega^2$. Also $\hat{\alpha}_{SIS}$ along $\beta$ decreases by increasing $b$ and increases with an increase in $\omega_0^2/\omega^2$, while $\hat{\alpha}_{SIS}$ along  $\omega_0^2/\omega^2$ decreases by increasing both $\beta \; \& \;b$.
 
 \subsection{Non-Singular Isothermal gas sphere}\label{A8.3}
We continue our investigation to analyze the motion of photons in a non-singular isothermal sphere ($NSIS$) plasma field, which serves as a suitable approximation for the physical analysis under consideration. Unlike the singular isothermal sphere ($SIS$) plasma, the $NSIS$ model is characterized by the absence of a singularity due to the presence of a definite core. The gas cloud sphere originates from this core, and its density distribution can be described as \cite{ad34,ad36}:
\begin{equation}\label{pa}
    \rho(r)=\frac{\sigma^2_v}{2\pi(r^2+r_c^2)}=\frac{\rho_0}{(1+\frac{r^2}{r_c^2})},\;\;\rho_0=\frac{\sigma^2_v}{2\pi r_c^2},
 \end{equation}
where $r_c$ denotes the core radius. Using Eq.(\ref{pl}), the concentration of plasma for $NSIS$ results in the following form:
\begin{equation}\label{pb}
    N(r)=\frac{\sigma^2_v}{2\pi k m_p (r^2+r_c^2)}.
 \end{equation}
Using Eqs.(\ref{pll}), (\ref{pa}) and (\ref{pb}), the plasma frequency $\omega_e$ yields
\begin{equation}\label{pc}
    \omega_e^2=\frac{K_e\sigma^2_v}{2\pi k m_p (r^2+r_c^2)}.
 \end{equation}
 The deviation of photons in the $NSIS$ plasma, in the gravitational lensing, produces a deflection angle which possesses all of its characteristics and can be computed for the SBRG BH as follows:
 \begin{eqnarray}\label{40}
    \hat{\alpha}_{NSIS}&=&\frac{1769472 \pi ^2 \sqrt{2} \beta  G^3 R_{s}^5 \omega_c^2}{175 b^9 r_{c}^2 \omega^2}+\frac{1769472 \pi ^3 \sqrt{2} \beta  G^3 R_{s}^3}{175 b^9}-\frac{1990656 \sqrt{2} \pi ^2 \beta  G^3 R_{s}^5 \omega_c^2}{175 b^7 r_{c}^4 \omega^2}+\frac{331776 \pi ^2 \sqrt{2} \beta  G^3 R_{s}^5 \omega_c^2}{25 b^5 r_{c}^6 \omega^2}\nonumber\\&-&\frac{82944 \sqrt{2} \pi ^2 \beta  G^3 R_{s}^5 \omega_c^2}{5 b^3 r_{c}^8 \omega^2}+\frac{12416 \pi ^3 \sqrt{2} b \beta  \sqrt{\frac{1}{b^2}} G^3 R_{s}^6 \omega_c^2}{r_{c}^{12} \omega^2}-\frac{6208 \sqrt{2} \pi ^3 \sqrt{\frac{1}{b^2}} \beta  G^3 R_{s}^6 \omega_c^2}{b r_{c}^{10} \omega^2}\nonumber\\&-&\frac{12416 \sqrt{2} \pi ^3 b \beta  G^3 R_{s}^6 \sqrt{\frac{1}{b^2+r_{c}^2}} \omega_c^2}{r_{c}^{12} \omega^2}+\frac{62208 \pi ^2 \sqrt{2} b \beta  G^3 R_{s}^5 \omega_c^2 \log \left(1-\frac{r_{c}}{\sqrt{b^2+r_{c}^2}}\right)}{5 r_{c}^{11} \omega^2 \sqrt{b^2+r_{c}^2}}\nonumber\\&-&\frac{62208 \sqrt{2} \pi ^2 b \beta  G^3 R_{s}^5 \omega_c^2 \log \left(\frac{r_{c}}{\sqrt{b^2+r_{c}^2}}+1\right)}{5 r_{c}^{11} \omega^2 \sqrt{b^2+r_{c}^2}}+\frac{b R_{s}^3 \omega_c^2 \log \left(\frac{r_{c}}{\sqrt{b^2+r_{c}^2}}+1\right)}{2 \pi  r_{c}^3 \omega^2 \sqrt{b^2+r_{c}^2}}-\frac{b R_{s}^3 \omega_c^2 \log \left(1-\frac{r_{c}}{\sqrt{b^2+r_{c}^2}}\right)}{2 \pi  r_{c}^3 \omega^2 \sqrt{b^2+r_{c}^2}}\nonumber\\&-&\frac{b R_{s}^2 \omega_c^2}{2 \omega^2 \left(b^2+r_{c}^2\right)^{3/2}}-\frac{6111 \pi ^3 \sqrt{\frac{1}{b^2}} \beta  G^3 R_{s}^6 \omega_c^2}{\sqrt{2} b^9 r_{c}^2 \omega^2}-\frac{6111 \pi ^4 \sqrt{\frac{1}{b^2}} \beta  G^3 R_{s}^4}{\sqrt{2} b^9}+\frac{3395 \pi ^3 \sqrt{2} \beta  \sqrt{\frac{1}{b^2}} G^3 R_{s}^6 \omega_c^2}{b^7 r_{c}^4 \omega^2}\nonumber\\&-&\frac{3880 \sqrt{2} \pi ^3 \sqrt{\frac{1}{b^2}} \beta  G^3 R_{s}^6 \omega_c^2}{b^5 r_{c}^6 \omega^2}+\frac{4656 \pi ^3 \sqrt{2} \beta  \sqrt{\frac{1}{b^2}} G^3 R_{s}^6 \omega_c^2}{b^3 r_{c}^8 \omega^2}+\frac{124416 \pi ^2 \sqrt{2} \beta  G^3 R_{s}^5 \omega_c^2}{5 b r_{c}^{10} \omega^2}-\frac{R_{s}^3 \omega_c^2}{\pi  b r_{c}^2 \omega^2}-\frac{R_{s}}{b}. 
 \end{eqnarray}
Graphical illustration of $\hat{\alpha}_{NSIS}$ is shown in Fig. \ref{plot:9a}. $\hat{\alpha}_{NSIS}$ behaves like $\hat{\alpha}_{uni}\;\&\;\hat{\alpha}_{SIS}$, moreover $\hat{\alpha}_{uni}>\hat{\alpha}_{SIS}>\hat{\alpha}_{NSIS}$ (see Fig. \ref{plot:10}).

\section{Magnification of gravitationally lensed image}\label{A9}
In this section, our focus is on examining the image magnification and brightness of the source in the presence of both uniform and non-uniform $SIS$ plasma fields. The distances between the relevant entities are denoted as follows: $D_s$ represents the distance from the source to the observer, $D_d$ represents the distance from the lens to the observer, and $D_{ds}$ represents the distance from the source to the lens. Additionally, we use $\delta$ and $\theta$ to denote the angular positions of the source and the image, respectively. Consequently, by considering small deflection angles, one can obtain the angular position or gravitational lensing using the following formula \cite{ad36,ad37,ad38}:
 \begin{equation}
    \theta D_{s}=\delta D_{s}+\hat{\alpha}D_{ds},
 \end{equation}
 which can also be written in terms of $\delta$ as
\begin{equation}
    \delta=\theta-\frac{D_{ds}}{D_{s}}\frac{\xi({\theta})}{D_{d}}\frac{1}{\theta},
 \end{equation}
where $\xi(\theta)=|\hat{\alpha}_{b}|b$ and $b = D_{d}\theta$ \cite{ad36}. The radius of Einstein's ring $R_s=D_d\theta_E$ refers to the radius of circular form of the image. The Einstein's angle $\theta_E$ between the source and the images in a vacuum is given by \cite{ad39} 
\begin{equation}
    \theta_E= \sqrt{2R_s\frac{D_{ds}}{D_dD_s}}.
 \end{equation}
The magnification of brightness can be mathematically expressed as \cite{ad34,ad39}
\begin{equation}
    \mu_\Sigma= \frac{I_{tot}}{I_{*}}=\sum_{K}\left|\left(\frac{\theta_k}{\delta}\right)\left(\frac{d\theta_k}{d\delta}\right)\right|,\;\;k=1,2,.....,j,
 \end{equation}
 where $I_{*},\;\&\;I_{tot}$ correspond to the notions of non-lensed brightness of the source and the total brightness of all the images, respectively. Magnification of the source is defined by \cite{ad34}
 \begin{eqnarray}
    \mu_{+}^{pl}&=& \frac{1}{4}\left(\frac{x}{\sqrt{x^2+4}}+\frac{\sqrt{x^2+4}}{x}+2\right),\label{b1}\\
\mu_{-}^{pl}&=& \frac{1}{4}\left(\frac{x}{\sqrt{x^2+4}}+\frac{\sqrt{x^2+4}}{x}-2\right),\label{b2}
 \end{eqnarray}
 where $x =\delta/\theta_0$ is a dimensionless entity \cite{ad36} and $\mu_{+}^{pl}$ and $\mu_{-}^{pl}$ correspond to the images in plasma field. We can derive an expression for the total magnification through Eqs.(\ref{b1}) and (\ref{b2}) as 
 \begin{eqnarray}
    \mu_{tot}^{pl}=\mu_{+}^{pl}+\mu_{-}^{pl}= \frac{x^2+2}{x\sqrt{x^2+4}}.
 \end{eqnarray}
Our objective is to investigate the brightness of the source and the image magnification in the plasma field surrounding a BH in SBRG. To achieve this, we examine the changes in the density distribution of the plasma field under two scenarios: (i) uniform plasma, and (ii) non-uniform plasma. These variations in the plasma density distribution will shed light on the impact of the plasma on the observed brightness and magnification of the images in the vicinity of the SBRG BH.
\subsection{Uniform Plasma}\label{A9.1} 
Here we consider the SBRG BH geometry to unveil the influence of the uniform plasma on the image magnification. The total magnification ($\mu_{tot}^{pl}$ and total deflection angel $\theta_{uni}^{ pl}$) can be calculated by the formula
\begin{equation}
    \mu_{tot}^{pl}=\mu_{+}^{pl}+\mu_{-}^{pl}=\frac{x_{uni}^2+2}{x_{uni}\sqrt{x_{uni}^2+4}},
 \end{equation}
 with
 \begin{eqnarray}
    \left(\mu_{+}^{pl}\right)_{uni}&=& \frac{1}{4}\Big[\frac{x_{uni}}{\sqrt{x_{uni}^2+4}}+\frac{\sqrt{x_{uni}^2+4}}{x}+2\Big],\label{b1}\\
\left(\mu_{-}^{pl}\right)_{uni}&=& \frac{1}{4}\Big[\frac{x_{uni}}{\sqrt{x_{uni}^2+4}}+\frac{\sqrt{x_{uni}^2+4}}{x_{uni}}-2\Big],\label{b2}
 \end{eqnarray}
 
\begin{equation}
    x_{uni}=\frac{\delta}{\left(\theta_{E}^{pl}\right)_{uni}},\;\; \theta^{pl}_{uni}=\theta _E \sqrt{\frac{b \alpha _b}{2 R_s}},
 \end{equation}
and
\begin{align}
    x_{uni}=x_0 \Big[\frac{b \left(\frac{196608 \pi ^3 \sqrt{2} \beta  G^3 R_{s}^3}{175 b^9}+\frac{\frac{1769472 \pi ^3 \sqrt{2} \beta  G^3 R_{s}^3}{175 b^9}-\frac{6111 \pi ^4 \sqrt{\frac{1}{b^2}} \beta  G^3 R^4}{\sqrt{2} b^9}-\frac{R_{s}}{b}}{1-B}-\frac{6111 \pi ^4 \sqrt{\frac{1}{b^2}} \beta  G^3 R^4}{10 \sqrt{2} b^9}-\frac{R_{s}}{b}\right)}{2 R_{s}}\Big]^{-\frac{1}{2}},
\end{align}
 where $x_0=\delta/\theta_E$.
\subsection{Non-uniform plasma}\label{A9.2}
 By employing the identical methodology as described earlier, it is possible to investigate the impact of the $SIS$ plasma on image magnification. This investigation yields the total magnification denoted as $\mu_{tot}^{pl}$ and the total deflection angle represented as $\theta_{SIS}^{ pl}$ for the $SIS$ plasma field, which can be determined using the following formula: 
 \begin{equation}
    \mu_{tot}^{pl}=\mu_{+}^{pl}+\mu_{-}^{pl}=\frac{x_{SIS}^2+2}{x_{SIS}\sqrt{x_{SIS}^2+4}},
 \end{equation}
 with
 \begin{eqnarray}
    \left(\mu_{+}^{pl}\right)_{SIS}&=& \frac{1}{4}\Big[\frac{x_{SIS}}{\sqrt{x_{SIS}^2+4}}+\frac{\sqrt{x_{SIS}^2+4}}{x}+2\Big],\label{b1}\\
\left(\mu_{-}^{pl}\right)_{SIS}&=& \frac{1}{4}\Big[\frac{x_{SIS}}{\sqrt{x_{SIS}^2+4}}+\frac{\sqrt{x_{SIS}^2+4}}{x_{SIS}}-2\Big],\label{b2}
 \end{eqnarray}
 
\begin{equation}
    x_{SIS}=\frac{\delta}{\left(\theta_{E}^{pl}\right)_{SIS}},\;\; \theta^{pl}_{uni}=\theta _E \sqrt{\frac{b \alpha _b}{2 R_s}},
 \end{equation}
and
\begin{eqnarray}
    x_{SIS}&=&x_0 \Big[\frac{b}{2 R_s} \Big[\frac{3538944 \pi ^2 \sqrt{2} \beta  B G^3 R_{s}^5}{385 b^{11}}+\frac{393216 \pi ^3 \sqrt{2} \beta  G^3 R_{s}^3}{35 b^9}-\frac{2 B R_{s}^3}{3 \pi  b^3}-\frac{22407 \pi ^3 \sqrt{\frac{1}{b^2}} \beta  B G^3 R_{s}^6}{4 \sqrt{2} b^{11}}\nonumber\\&-&\frac{67221 \pi ^4 \sqrt{\frac{1}{b^2}} \beta  G^3 R_{s}^4}{10 \sqrt{2} b^9}-\frac{2 R_{s}}{b}\Big]\Big]^{-\frac{1}{2}},
\end{eqnarray}
We plot the total image magnification in uniform and $SIS$ plasma in Figs.~\ref{plot:11} and \ref{plot:12}. We find that magnification increases in a higher concentration of plasma field. It is interesting to notice that the image magnification in uniform plasma is much higher as compared to the $SIS$ plasma field.
\section{Effective potential and effective force}

\begin{figure}
    \includegraphics[scale=0.3]{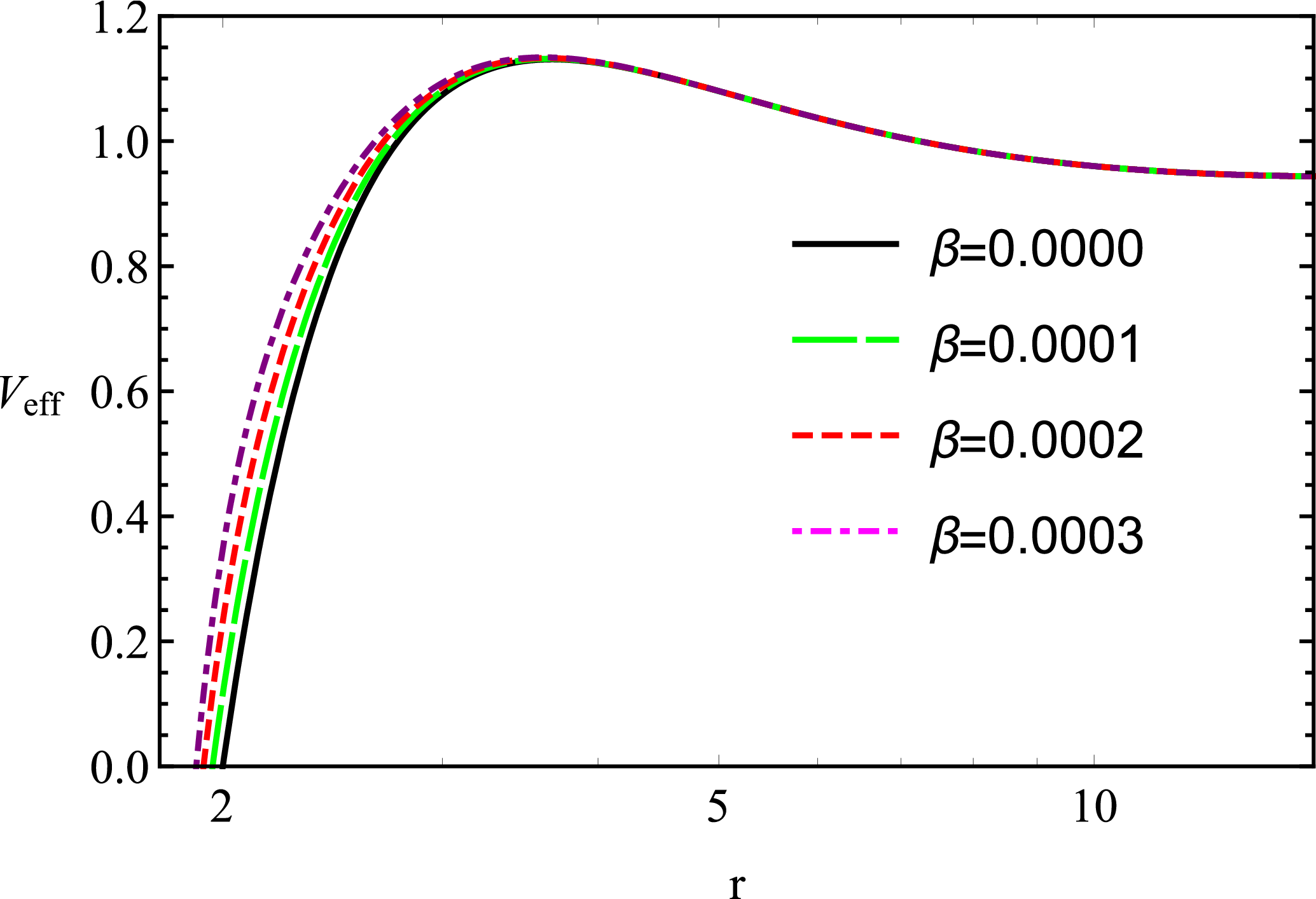}
    \caption{Effective potential with fixed values $G=1,\;M=1$.}
    \label{plot:13}
    \end{figure}

The equation of geodesics depending upon the radial is defined as:
\begin{equation}\label{12}
    \left(\frac{dr}{d\tau}\right)^2=\mathcal{E}^2-V_{eff}(r)=\mathcal{E}^2-f(r)\left(1+\frac{\mathcal{L}^2}{r^2}\right),
\end{equation}
where
\begin{equation}\label{13}
    V_{eff}(r)=f(r)\left(1+\frac{\mathcal{L}^2}{r^2}\right).
\end{equation}
Here, $V_{eff}(r),\;\epsilon,\;\&\;\mathcal{L}$ represent the effective potentials for the test particles of radial motion, geodesic motion, and angular momentum, respectively. The Fig. (\ref{plot:13}) shows the radial propagation of $V_{eff}$ for the current analysis. Interestingly, $V_{eff}$ is increasing within the scope of the involved parameter $\beta$.
\begin{figure}
\includegraphics[scale=0.24]{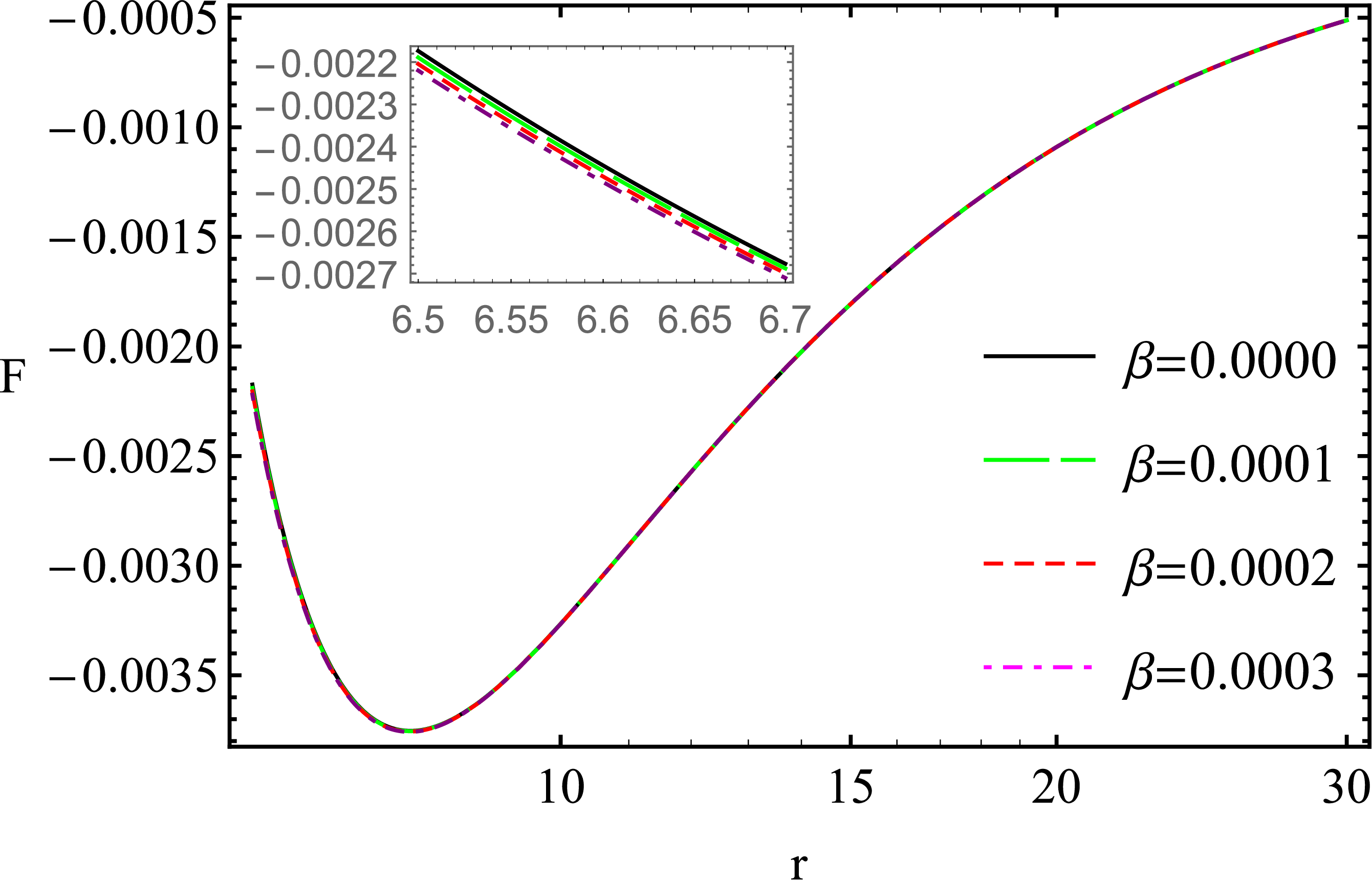}
\caption{Effective force with fixed values $G=1,\;M=1$.}
\label{plot:14}
\end{figure}
The behavior of effective force can be used to infer whether the test particle is moving toward or away from the central source. The following formula can be used to calculate the effective force felt by a test particle in the gravitational source field \cite{153}: 
\begin{equation}
    F=-\frac{1}{2}\frac{\partial\; V_{eff}(r)}{\partial r}.
\end{equation}
In the case of SBR geometry, the computed effective force expression is as follows:
\begin{eqnarray}
F&=&\frac{-12 G^3 M^3+8 G^2 M^2 r-G M r^2}{2 r^2 (3 G M-r)^2}+\frac{1024 \beta}{5 r^{11} (3 G M-r)^3}  \Big[34920 \pi ^3 \sqrt{2} G^{10} M^7-65214 \sqrt{2} \pi ^3 G^9 M^6 r\nonumber\\&+&45742 \pi ^3 \sqrt{2} G^8 M^5 r^2-14329 \sqrt{2} \pi ^3 G^7 M^4 r^3+1701 \pi ^3 \sqrt{2} G^6 M^3 r^4\Big].
\end{eqnarray}
A more accurate interpretation of the effective force's influence can be obtained from the graphical analysis shown in Fig.~\ref{plot:13}. Along the radial motion, one can observe how the effective force fluctuates, going from attractive to repulsive and back to attractive.

\section{CONCLUSIONS}\label{VI}
In the current manuscript, we have discussed thermodynamics and lensing with plasma of Schwarzschild type BHs in SBR modified theory of gravity. For the current analysis, we have investigated the thermal stability of the solution by analyzing its temperature and the heat capacity and mass evaporation through emission energy. The positive behavior of the metric function $f(r)$ and mass $M$ ensure the physical existence of BH in SBR modified theory of gravity. Besides, we have also calculated weak gravitational lensing in the presence of plasma and its magnification in the background of uniform and non-uniform  plasma. Some important insights regarding the current study are listed as: 
\begin{itemize}
  \item The lapse function $f(r)$ is plotted in Fig. (\ref{p1}), from this figure, one can confirm the singularity. The positive event horizon with singularity can be confirmed from the same figure for the different positive values of involved parameter$\beta$. It is necessary to mention that the horizon radius, of the Schwarzschild-like BH in the SBRG increase with an increase in the parameter $\beta$. It should be noted the horizon radius is of the Shwarzschild-like BH in the SBRG is smaller than that of the Schwarzschild BH in the Einstein theory of gravity \cite{adn47}.
  
  \item The graphical behavior of the emission energy has provided in Fig. (\ref{plot:6}). One can see that $\varepsilon_{\omega t}$ increases by decreasing values of important parameter $\beta$.
 
  \item We have plotted $S_C$ by considering different choices of the correction parameter: $\gamma=0$. The plot, which has shown in Fig. (\ref{plot:5}) demonstrate the entropy of spacetime consistently increases across the entire range considered for different values of $\gamma$ with positive fixed values of $\beta$. Moreover, fluctuation can be observed for nonzero values of the parameter $\beta$. 
  \item The graphical behavior of the deflection angle has provided in Fig.~(\ref{plot:7}). It has noted that  $\hat{\alpha}_{uni}$ along impact parameter $b$ decrease by increase in $\beta$ and increase with an increase in $\omega_0^2/\omega^2$, $\hat{\alpha}_{uni}$ along $\beta$ decrease by increase $b$ and increase with an increase in $\omega_0^2/\omega^2$, while $\hat{\alpha}_{uni}$ along  $\omega_0^2/\omega^2$ decrease by increase both $\beta \; \& \;b$.
   \item For the $SIS$ plasma graphical behavior of the deflection angle has presented in Fig.~(\ref{plot:8}). 
    \item Further, the behavior of $\hat{\alpha}_{NSIS}$ has shown in Fig. (\ref{plot:9a}). The $\hat{\alpha}_{NSIS}$ behaves like $\hat{\alpha}_{uni}\;\&\;\hat{\alpha}_{SIS}$, moreover the $\hat{\alpha}_{uni}>\hat{\alpha}_{SIS}>\hat{\alpha}_{NSIS}$, one can see from the Fig. \ref{plot:10}).
  \item The magnification images in uniform and $SIS$ plasma have presented in Fig.~(\ref{plot:11}) and Fig.(\ref{plot:12}) respectively. It is interesting to notice that the image magnification in uniform plasma is much higher as compared to the $SIS$ plasma field.

  \item The effective potential for the current analysis is observed for four different values of involved parameter$\beta$. The positive behavior of effective potential including minimum and maximum behavior, can be confirmed from the Fig. (\ref{plot:13}). 
  \item The Fig. (\ref{plot:14}) shows the graphical representation of effective force.  Along the radial motion, one can observe how the effective force fluctuates, going from attractive to repulsive and back to attractive. It can be confirmed from the Fig. (\ref{plot:14}) that effective force remain negative throughout the configuration, which is necessary features for the stable orbit.
\end{itemize}
 Overall, our calculated results in the current analysis are physical viable and have a good agreement with the previous published results in the literature.    

 \section*{Acknowledgments}
This work is partly supported by Grants F-FA-2021-432, F-FA-2021-510, and MRB-2021-527 of the Ministry of Higher Education, Science and Innovations of the Republic of Uzbekistan.


\end{document}